\documentclass[prl,showpacs,twocolumn,a4paper,amsmath,floatfix]{revtex4-1}
\pdfoutput=1

\usepackage{graphicx}
\usepackage{dcolumn}
\usepackage{bm}
\usepackage{color,soul}
\usepackage{amsmath}

\begin{document}

\title{Determination of the positions and orientations of concentrated rod-like colloids from 3D microscopy data}

\author{T. H. Besseling} \thanks{Both authors contributed equally}
\affiliation{Soft Condensed Matter, Debye Institute for
NanoMaterials Science, Utrecht University, Princetonplein 1,
NL-3584 CC Utrecht, the Netherlands}

\author{M. Hermes} \thanks{Both authors contributed equally}
\affiliation{Soft Condensed Matter, Debye Institute for
NanoMaterials Science, Utrecht University, Princetonplein 1,
NL-3584 CC Utrecht, the Netherlands}

\author{A. Kuijk}
\affiliation{Soft Condensed Matter, Debye Institute for
NanoMaterials Science, Utrecht University, Princetonplein 1,
NL-3584 CC Utrecht, the Netherlands}

\author{B. de Nijs}
\affiliation{Soft Condensed Matter, Debye Institute for
NanoMaterials Science, Utrecht University, Princetonplein 1,
NL-3584 CC Utrecht, the Netherlands}

\author{T.-S. Deng}
\affiliation{Soft Condensed Matter, Debye Institute for
NanoMaterials Science, Utrecht University, Princetonplein 1,
NL-3584 CC Utrecht, the Netherlands}

\author{M. Dijkstra}
\affiliation{Soft Condensed Matter, Debye Institute for
NanoMaterials Science, Utrecht University, Princetonplein 1,
NL-3584 CC Utrecht, the Netherlands}

\author{A. Imhof}
\affiliation{Soft Condensed Matter, Debye Institute for
NanoMaterials Science, Utrecht University, Princetonplein 1,
NL-3584 CC Utrecht, the Netherlands}

\author{A. van Blaaderen}
\email{A.vanBlaaderen@uu.nl}\homepage{www.colloid.nl}
\affiliation{Soft Condensed Matter, Debye Institute for
NanoMaterials Science, Utrecht University, Princetonplein 1,
NL-3584 CC Utrecht, the Netherlands}

\date{\today}

\begin{abstract}
Confocal microscopy in combination with real-space particle tracking has proven to be a powerful tool  in scientific fields such as soft matter physics, materials science and cell biology. However, 3D tracking of anisotropic particles in concentrated phases remains not as optimized compared to algorithms for spherical particles. To address this problem, we developed a new particle-fitting algorithm that can extract the positions and orientations of fluorescent rod-like particles from three dimensional confocal microscopy data stacks, even when the fluorescent signals of the particles overlap considerably. We demonstrate that our algorithm correctly identifies all five coordinates of  uniaxial particles in both a concentrated disordered phase and a liquid-crystalline smectic-B phase. Apart from confocal microscopy images, we also demonstrate that the algorithm can be used to identify nanorods in 3D electron tomography reconstructions. Lastly, we determined the accuracy of the algorithm using both simulated and experimental confocal microscopy data-stacks of diffusing silica rods in a dilute suspension. This novel particle-fitting algorithm allows for the study of structure and dynamics in both dilute and dense liquid-crystalline phases (such as nematic, smectic and crystalline phases) as well as the study of the glass transition of rod-like  particles in three dimensions on the single particle level.
\end{abstract}

\maketitle

\section{Introduction}

Colloidal particles are applied throughout industry, for example in paints, personal care products, food, ceramics and pharmaceutics \cite{Croll2002,Letchford2007,Sozer2009}. Additionally, they are also applied in recent commercially available products such as the electronic ink in e-readers \cite{Sucher1998}. As a result, the characterisation of the structure and dynamics of
colloidal suspensions is important for many industrial applications.
Furthermore, hard-sphere colloidal suspensions have proven to serve as a 
model system to investigate phenomena such as crystallization, the glass
transition and flow induced behaviour on the single particle level
\cite{VanBlaaderen1995,Weeks2000,Dassanayake2000,Gasser2001,Dinsmore2001,Cohen2006,Yethiraj2003,Besseling2012a}. 
In many of these studies, an image processing technique was applied based on 
 the algorithm described by Crocker and Grier \cite{Crocker1996}. In their algorithm, spherical particles are
located in 2D digital microscopy images using a local brightness maxima criterion. The position is refined by calculating the brightness-weighted centroid of a cluster of pixels. This
method was extended to 3D either slice-by-slice
\cite{VanBlaaderen1995,Dassanayake2000} or by considering the full 3D image
\cite{Dinsmore2001}. Crocker and Grier also reported a method to obtain the trajectories of the individual particles in time, known as particle tracking
\cite{Crocker1996}. Since then, there have been numerous algorithms that locate or track spherical particles with increased accuracy or performance \cite{Lu2007,Jenkins2008,Gao2009,Besseling2009a,Vissers2011,Kurita2012a,Leocmach2013}. These extensions and alternatives are all based on processing images of \textit{spherical} particles. However, due to recent progress in particle synthesis, well-defined (shape) anisotropic colloids are becoming widely available, see e.g.\ Refs.~\citenum{Kuijk2011,Kuijk2012a,Mohraz2005,Rossi2011,Elsesser2011,Peng2012}. These particles can often be observed directly with a (confocal) microscope and therefore enable quantitative measurement of not only their positional but also their rotational degrees of freedom. Therefore, a rapid increase in the number of algorithms that extract coordinates of anisotropic particles from microscopy images has taken place \cite{Mohraz2005,Anthony2006,Han2006,Hunter2011,Zhao2011,Chakrabarty2013}. Most of these algorithms are based on processing of 2D (bright-field) images of quasi-2D systems. Mohraz and Solomon, however, were one of the first to determine the 3D position and orientation of uniaxial ellipsoidal particles, i.e.\ all five coordinates, using 3D confocal microscopy and a novel anisotropic feature-finding algorithm \cite{Mohraz2005}. Their algorithm identifies the points that are located on the central axis (or backbone) of a rod. These points are then grouped together by cluster analysis as individual rod-backbones, from which the centroid location and orientation are determined. This algorithm enabled the quantitative determination of the 3D translational and rotational motion of a dilute suspension of ellipsoids
\cite{Mukhija2007}. 
\\ \indent Quantitative 3D real-space study of concentrated phases of anisotropic particles is, however, much less progressed compared to studies on spherical colloids. Progress has been made for suspensions of
ellipsoids, where nematic order was found using a centrifugal field
\cite{Mukhija2011} and local crystalline order with an external electric field
\cite{Shah2012}. In contrast with the system of ellipsoidal particles, it was recently shown by some of us that a  system of fluorescent silica rod-like particles forms both nematic and smectic phases in equilibrium  \cite{Kuijk2011,Kuijk2012a}. However, determination of all the 3D positions and orientations of the particles in these dense phases was not possible due to the significant overlap of fluorescent signals. 
In this paper we demonstrate a novel 3D image processing algorithm that is capable of quantifying fluorescent silica rods in concentrated (liquid-crystalline) phases. The algorithm is tailored to work  even when the fluorescent signals of the particles overlap considerably and a threshold method and subsequent clusters analysis alone does not suffice. 
 The algorithm in principle also works for other uniaxial particles such as ellipsoids or dumbbells.

This paper is organized as follows. First, we describe the basics of
particle-locating algorithms. Second, we describe our algorithm in detail.
Third, we demonstrate the performance of the algorithm with 3D image stacks of
concentrated fluorescent silica rods. Then, we illustrate that our algorithm can also be applied to 3D electron tomography data of gold nanorods. Next, we evaluate the accuracy of the algorithm by measuring the translational and rotational motion of non-overlapping rod-like particles. Finally, we compare  our results with recent progress in the field and give an outlook on further studies that the algorithm enables.
\section{Methods}
\subsection{Locating particles in confocal microscopy data sets}
The aim is to identify and locate (rod-shaped) particles in a set of real-space images (or snapshots) 
and to obtain the full configuration of the system.
A specific configuration of a system of particles is given by a set of parameters, one  
for each degree of freedom of every particle.
In the case of rods, these degrees of freedom for particle $i$
are centre position $\mathbf{r}_i$, orientation $\mathbf{\hat{u}}_i$ and possibly length $l_i$, diameter $d_i$ and brightness $b_i$. If the length and diameter are known in advance they can be fixed, but if the
particles vary in size they can also be left as free parameters.
If the particles vary in brightness this can be added as an additional degree of freedom.
Variations in brightness can be caused by the synthesis method, scattering or shading in the sample, but also by photo bleaching. In the case of fully symmetric, homogeneously dyed rods it is not possible to distinguish between the two ends of the rods. However, we also synthesised rods with a gradient in brightness,
with one bright and one much darker end \cite{Kuijk2014a}, of which the orientation could be fully determined. 
To keep the notation short we introduce ${\mathbf p}_i=\{ {\mathbf{\hat u}}_i,l_i,d_i,b_i\}$ which contains 
all the degrees of freedom except the position.

To obtain the configuration (${\mathbf r}_i$ and ${\mathbf p}_i$) we need to elaborate on what is measured. 
In case of fluorescent confocal laser scanning microscopy we can assume that
the imaging system is linear so that we can add intensities. The measured image
intensity $M(\mathbf{r})$ at position $\mathbf{r}$ can be written as
the sum of the ideal (noiseless or averaged) images of the single particles,
\begin{equation}
M(\mathbf{r}) = \sum_{i=1}^N \mathrm{RSP}(\mathbf{r}-\mathbf{r}_i,\mathbf{p}_i),
\end{equation}
and $\mathrm{RSP}(\mathbf{r},\mathbf{p}_i)$ is the image of a single particle placed in the origin,
 or rod spread function (RSP).
The image of a single particle at the origin depends on all the internal degrees of
freedom of the particle such as orientation, length, diameter and brightness, but also on the
point spread function (PSF) of the imaging system. It is given by
\begin{eqnarray}
\mathrm{RSP}(\mathbf{r},\mathbf{p}_i) &=&
\int \mathrm{d}\mathbf{r}' \rho_\mathrm{dye}(\mathbf{r}',\mathbf{p}_i) \mathrm{PSF}(\mathbf{r}-\mathbf{r}')\nonumber\\
&=&(\rho_\mathrm{dye}(\mathbf{p}_i) \ast \mathrm{PSF})(\mathbf{r}),
\end{eqnarray}
which is a convolution ($\ast$) of the dye distribution $\rho_\mathrm{dye}(\mathbf{r},\mathbf{p}_i)$ 
of particle $i$ placed in the origin and the PSF. In a dilute sample this $\mathrm{RSP}(\mathbf{r},\mathbf{p}_i)$ can be
measured directly but it can also be calculated when the dye distribution is simple
and the parameters of the optical systems are known.

Different approaches to obtain the particle coordinates are possible. If all
the parameters such as the PSF and RSP are known, the locating problem becomes
in principle a deconvolution. However, the RSP and the PSF can be time consuming
to determine accurately, and deconvolutions are sensitive to small changes in
the kernel function \cite{Cannell2006}. This is unfortunate since e.g.\ polydispersity will introduce
changes in the RSP which would make the deconvolution difficult.
If the RSP is not known, there exist several other possible options.
The first option is to assume that the overlap between the RSPs is
not too severe and to determine centre-of-mass and orientation with methods that are
insensitive to the details of the optical system. This is the method used 
by centroiding algorithms and is also the method used in this article.

Another option is to use a Bayesian method \cite{Besag1991}. This method searches for
the configuration that has the largest probability of having resulted in the observed
image. This method has proven to work well for two-dimensional data sets \cite{Al-Awadhi2004}. It is, however,
slow and complex and therefore not practical for large three-dimensional data sets.

\subsection{Generation of test images}
To test our algorithm we generated 8-bit confocal-like images from sets of computer-generated particle trajectories. Using the centres-of-mass ${\mathbf r}_i$ and particle
orientations ${\mathbf{\hat u}}_i$, we generated 3D stacks of $xy$-images of spherocylinders with aspect ratio $l$/$d$ = 5, where $l$ is the end-to-end length of the particle and $d$ the diameter.
This was done by calculating the closest distances $D$ to a line segment, representing the backbone of a particle.
The distance from a point in the origin to a line segment from $\mathbf{x}_1$ to $\mathbf{x}_2$ with length $l= |\mathbf{x}_1-\mathbf{x}_2|$
is given by
\begin{equation}
	D(\mathbf{x}_1,\mathbf{x}_2)=\left.
		\begin{array}{rcl}
			|\mathbf{x}_1|            & \text{if} &\alpha<0,			\\
			\sqrt{|\mathbf{x}_1|^2-\alpha^2 } &\text{if} &0<\alpha<l,	\\
			|\mathbf{x}_2|            & \text{if} &\alpha>l,			\\
		\end{array} \right.
\end{equation}
where
$\alpha=(\mathbf{\hat{u}} \cdot \mathbf{x}_1)$ and $\mathbf{\hat{u}}= (\mathbf{x}_1-\mathbf{x}_2)/l$ the unit vector along the length of the line segment. If this distance was less than the diameter of the particle, the pixel was given a value of 0.95. This was then repeated for all particles. We approximated the effect of the PSF in our test images by convolving them with a Gaussian kernel with fixed standard deviation $\sigma_x/d = \sigma_y/d = 0.3$ and $\sigma_z/d = 0.3,0.6$ and $0.9$ with $d$ the diameter of the particle. The full-width-at-half-maximum (FWHM) of the Gaussian function, given by  $2\sqrt{2 \ln 2}\,\sigma_i$, is a direct measure of the resolution of the images. Besides variation of resolution, we also varied the amount of noise in the images. Although noise from modern detectors is essentially photon-limited, suggesting a Poisson distribution \cite{Art2006}, we added noise to each pixel in our images with a simple Gaussian distribution with standard deviation $\sigma_n = 0.10 - 0.30$. Because the amount of noise is known a priori, it is still straightforward to calculate the signal
to noise ratio (SNR), which we define as SNR = $(\sigma_g^2 / \sigma_n^2 - 1)^{1/2}$, with $\sigma_g^2$ the variance of the constructed image and $\sigma_n^2$ the variance of the noise \cite{Jenkins2008}. Finally, we converted all our data, with pixel-values between 0 and 1, to 8-bit grayscale tiff images. 

\subsection{Our algorithm}

\begin{figure}[h!]
\includegraphics[width=0.45\textwidth]{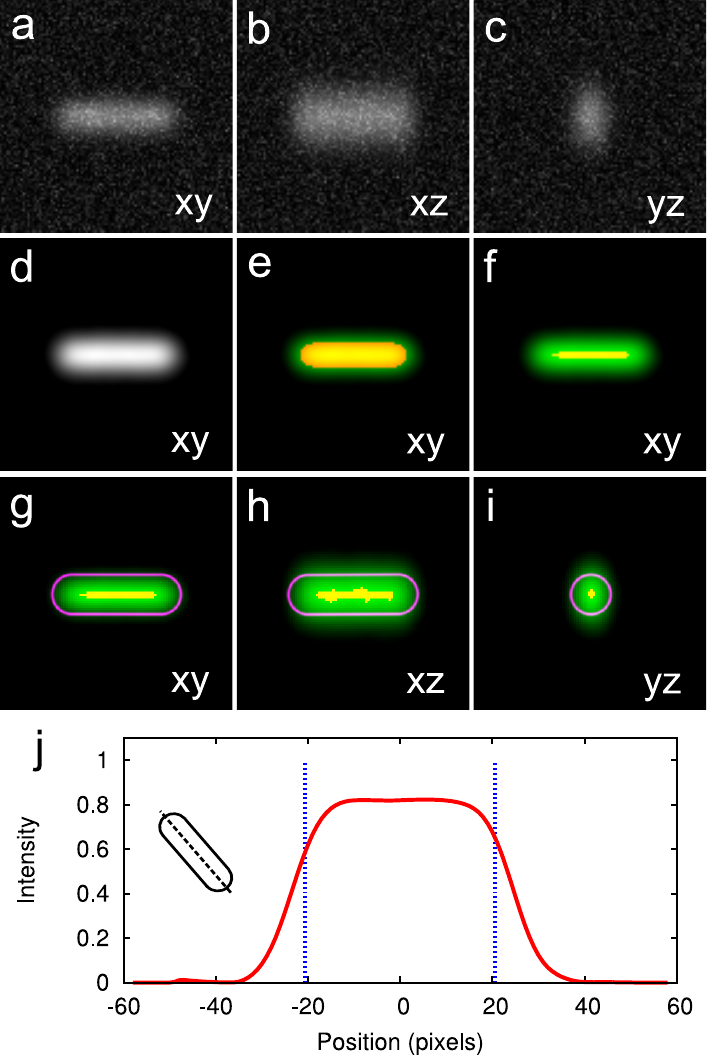}
\caption{
\label{fig:single}
The different stages of identification of the position and orientation of a
single rod-shaped particle.
(a,b,c) Orthogonal slices through a computer generated 3D image stack. The particle has a diameter
of 13.0 pixels and is blurred with a Gaussian kernel with width $\sigma_x/d = \sigma_y/d  = 0.3$, and $ \sigma_z/d  = 0.9$. Pixel noise has been added by adding Gaussian noise with $\sigma_n$ = 0.1.
(d) The same image after the filter-step. 
(e) After a threshold step, the pixels above the threshold are marked in yellow.
(f) After a backbone step, the pixels identified as backbone pixels are marked
in yellow.
(g,h,i) The rod as it is located, viewed from the $xy$, $yz$ and $xz$ plane.
(j) The histogram of the average intensity along the rod length after smoothing and background removal.
The dashed vertical lines mark the fitted end-points of the rod.
}
\end{figure}

To demonstrate the three-dimensional rod tracking algorithm we will first illustrate all
steps of the algorithm with an artificially created set of images of a single rod, shown in  Fig.~\ref{fig:single}. This will allow us to demonstrate clearly what is going on on a single pixel/voxel level. Later we will demonstrate how the algorithm fares with real colloidal suspensions. The following description is for three dimensions but most of the steps are straightforward to modify for two dimensions. 

\paragraph{Reading}
In Figs.~\ref{fig:single}a-c we show three orthogonal slices through a generated 3D image that acts as the source image. The particle shown in the image has a diameter $d = 13.0$ pixels, and is blurred with a Gaussian kernel $\sigma_x/d = \sigma_y/d = 0.3$ and $\sigma_z/d$ = $0.9$. Gaussian pixel noise of $\sigma_n = 0.1$ was added to the image. 
The first step is to read in these source images. To avoid accumulating rounding errors
and to allow the use of images of arbitrary bit depth we perform all image manipulations
on floating point numbers between zero and one. 
Next, the image is rescaled to make sure the voxels are cubic, which is often not the case for confocal microscopy image stacks. The rescaling avoids having to account for different $x$, $y$ and $z$ scales in all following routines. To make sure no information is lost, this is done by enlarging the image using a bicubic interpolation.
Care should be taken not to use overexposed images since this will result in a loss of information and an increase of positional error. See Ref.~\citenum{Jenkins2008} for a more detailed description and the optimal shape of the intensity histogram.
We generally choose the magnification such that the particles are approximately 10 pixels in diameter. Larger magnification results in a large file size without any additional benefit.

\paragraph{Filter}
The aim of the first filter step is to reduce image noise. We apply a Gaussian blur to the image, i.e. a convolution with a Gaussian kernel, that acts as a low pass filter. The optimal width of the function depends on the noise level in the images; a value between 1.5 and 3 pixels was found to give the best results for the images obtained in the present paper. A value that is too large will result in the loss of resolution and in missing particles, a value that is too small will result in additional, incorrectly identified, particles. 
To ensure a black background for the particles, a background value 
is subtracted from every pixel. This background value is assumed to be mostly the result of photon noise, but it can also originate from other sources such as fluorescence from the solvent or immersion fluid. Pixels that have a negative value after the background value has been subtracted, are set to zero.
In most cases a background value between 0.01 and 0.1 is used. This value should be chosen such that
approximately half the empty pixels (not containing a particle) of the image are zero.
We also save a copy of the image that has not been filtered. This allows us to perform
the final fitting step on the original image. An example of a computer-generated image that has been filtered is shown Fig.~\ref{fig:single}d.

\paragraph{Well separated particles}
When the intensity distributions of the individual particles do not overlap significantly we apply what we call a threshold method. This threshold method works as follows. A typical value for the threshold is between 0.4 and 0.7 and can be determined by plotting a histogram or by a quick test on a single image in a program like Photoshop, Gimp or ImageJ. The next step is to group all connected pixels
above the threshold value into sets, as described in the next section. This methods works when these sets of pixels belong each to a single particle and each particle only corresponds to a single set
of pixels. In Fig.~\ref{fig:single}e an example is shown of the threshold method applied to a single particle. All pixels above the threshold are marked in yellow. 
The particle coordinates can be obtained by applying a fit to these sets of pixels, as described later in this section. When this threshold method works, it is preferred
over more complex methods since it is both robust and accurate. 

\paragraph{General case}
When a threshold does not successfully separate the image into regions belonging to
single particles another method has to be used. The first step of this method is similar to the Crocker and Grier algorithm and is aimed at providing the final
fitting steps with a good initial starting point.

In this step, we roughly locate the line segment starting from one end of the rod and ending at the other end, called the backbone of the particle. To locate the backbone, we look at all voxels brighter than a predetermined cut-off value. A good value for this is in general between 0.1 and 0.5 depending on the intensity fluctuations
between the rods. For these bright pixels we then check whether they are part of a backbone. To do this
we first note that all local maxima should be part of the backbone.
To check if the brightness of the pixel is a local maximum we
compare its intensity to that of all pixels within a distance $r_\mathrm{bb}$. 
If none of these pixels are brighter the pixel is a local maximum.
To find the parts of the backbone that are not on a local maximum, we look at the distribution
of brighter pixels around the pixel in question. If the pixel is part of the backbone
they should be on a ridge. Backbone pixels can have brighter pixels to one side or two sides
but all these brighter pixels should be more or less on a line through the pixel in question.
So to check if the pixel is part of a backbone we need to check if the pixels
brighter than the pixel in question are on a straight line.
To do this we fit a line to these bright pixels and sum the squared residuals $\chi$, the
squared distance between the brighter pixels and the line. If these bright pixels
are part of the backbone of a rod this number will be low since the pixels
will form an almost perfect line while on other places they will not form a line
and the residuals will be much higher. We found that $r_\mathrm{bb}=3$
pixels and a maximum value $\chi_\mathrm{max}=80$ work well for all our data.
This step depends on the initial filtering and on the thickness of the rod in pixels.
Fig.~\ref{fig:single}f shows the pixels that have been identified as backbone pixels
in yellow.

After having identified the backbone pixels, we group them into connected clusters.
Due to noise there can be small gaps between the backbone pixels of a rod, so we use the same
search range $r_\mathrm{bb}$ as before to identify neighbouring pixels. This should work
as long as the diameter of a rod is larger than $r_\mathrm{bb}$.

We now have groups of pixels most likely belonging to a single rod. To continue, we fit
(least square) a straight line to these pixels using a singular value decomposition
and the algorithm described in Ref.~\citenum{Shakarji1998}. The coordinates resulting from this fit are accurate, but still have a strong pixel bias since they only fit to a few backbone pixels. To eliminate this bias and
to obtain more accurate results, we use these coordinates, lengths and orientations as
a starting point to fit the real image again.

\paragraph{Fitting}
The fitting steps work best when applied to the unfiltered image. The Gaussian blur filter 
will result in an additional overlap of the RSPs which can result in a decreased accuracy.
The fitting is done in three steps; first the
centre of mass of each group of pixels is computed, then the orientation is fitted and finally the length is
fitted. The position is taken from the centre of mass, weighted with the pixel intensity, of the
pixels within half a diameter from the previous fit. The orientation is obtained by
fitting a straight line to these pixels where the fit is weighted with the intensity
of the pixels using the same least square fitting algorithm as for the backbones.
The length is obtained by calculating the average intensity of pixels along the rod
length, see Fig.~\ref{fig:single}j. The histogram that is obtained from this is smoothed with a Gaussian kernel to avoid noise. The end points are then obtained by determining where the histogram value
drops below $I_\mathrm{end}I_\mathrm{max}$, where $I_\mathrm{max}$ is
the maximum intensity value in the smoothed histogram and $I_\mathrm{end}$
is a parameter that can be set manually. Usually a value of $I_\mathrm{end}$ = 0.6 - 0.8 was found to give good results, see the (blue) dashed lines in Fig.~\ref{fig:single}j. To obtain sub-pixel accuracy we fit a
straight line to the 2 pixels above and 2 pixels below the point where the histogram
crosses this value. To determine which pixels to take into account in the generation of the histogram
and the other fits, we use the pixels within one radius of the central line segment
of the previous fit. Therefore, the result of the fit might improve when
the step is repeated. The fitting algorithm normally converges in one or two steps. If this is not the case there is something wrong with the data or one of the parameters. Figs.~\ref{fig:single}g-i show the same orthogonal sections as Figs.~\ref{fig:single}a-c with the backbone of the rod highlighted in yellow and the outline of the rod (resulting from the fit) highlighted in magenta.

\paragraph{Filtering}
The final step is to filter out particles that are found more than once, particles
that do not contain enough intensity or sometimes particles that are not long enough.
Ideally not much filtering is required.

\subsection{3D particle tracking}
To study particle dynamics, we applied our algorithm to time-series of 3D image-stacks. We first identified the positions and orientations of the rods in each 3D stack
separately. Then, we obtained the particle
trajectories using standard IDL-based routines
\cite{Crocker1996}. To uniquely track the tip of the (up-down
indistinguishable) rods, it is required that the angular displacements between
successive frames $[{\mathbf{\hat u}}(t+1) - {\mathbf{\hat u}}(t)]^2 < 2$. Therefore, care was
taken that displacements with $[{\mathbf{\hat u}}(t+1) - {\mathbf{\hat u}}(t)]^2 > 2$ were
negligible. We then calculated the mean squared displacement (MSD) and the mean
squared angular displacement (MSAD). We fitted the MSD to the expression 
\begin{equation}\label{eq:msd}
 \langle \Delta {\bf r}^2 (t) \rangle =6\,D_t\,t + 6 \, \epsilon_t^2, 
\end{equation}
with $D_t$ the rotationally averaged translational diffusion coefficient and
$\epsilon_t$ the error in measurement of each of the coordinates of the particle \cite{Savin2005}. 
For the MSAD we used the expression
\cite{Dhont1996,Cheong2010}
\begin{equation}\label{eq:msad}
 \langle \Delta {\mathbf{\hat u}}^2 (t) \rangle = 2[1-(1-\epsilon_r^2)\exp({-2 D_r t})], 
\end{equation}
with $D_r$ the average rotational diffusion coefficient and $\epsilon_r$ the measurement error in the determination of ${\mathbf{\hat u}}(t)$.
For short times, equation \eqref{eq:msad} reduces to
\begin{equation}
\langle \Delta {\mathbf{\hat u}}^2 (t) \rangle = 4D_r t + 2\epsilon_r^2\,.
\end{equation}\label{eq:msad_short}
To estimate the sedimentation velocity at infinite dilution,
assuming complete decoupling of rotations, translations and sedimentation \cite{Homogeneous1979}, we
use the Svedberg equation \cite{Svedberg1940}
\begin{equation}\label{eq:sed}
 v_{sed} = \frac{v_p\,D_t\,g\,(\rho_p-\rho_s)}{k_B\,T}, 
\end{equation}
with  $v_p$ the volume of the particle, $g$ the gravitational
acceleration, $\rho_p$ the mass density of the particle and $\rho_s$ the mass density of
the solvent.


\subsection{Expressions for the diffusion coefficients}
To test the validity of our experimental measurements of the diffusion coefficients, we compared them to  analytical expressions for hard cylinders at infinite dilution, as proposed by Tirado, Martinez and de la Torre \cite{Tirado1984},
\begin{align}
D_\perp &= \frac{k_B T}{4 \pi \eta \,l}(\log p + \delta_\perp), \\
D_\parallel &= \frac{k_B T}{2 \pi \eta \,l}(\log p + \delta_\parallel), \\
D_t &= \frac{2}{3}D_\perp + \frac{1}{3}D_\parallel \label{eq:dtrans}, \\ 
D_r &= \frac{3 k_B T}{\pi \eta \,l^3}(\log p + \delta_r) \label{eq:Dr},
\end{align}
with $\eta$ the solvent viscosity, $p = l/d$ the aspect ratio of the particle and $\delta_i$ a correction term for the finite aspect ratio of the cylinders, given by \cite{Tirado1984}
\begin{align}
\delta_\perp &= 0.839 + 0.185/p + 0.233/p^2, \\
\delta_\parallel &= - 0.207 + 0.980/p - 0.133/p^2, \\
\delta_r &= - 0.662 + 0.917/p - 0.050/p^2.
\end{align}

\subsection{Experimental methods} 
\subsubsection{Dense sediments of silica rods} 
\begin{figure}[ht!]
\includegraphics[width=0.45\textwidth]{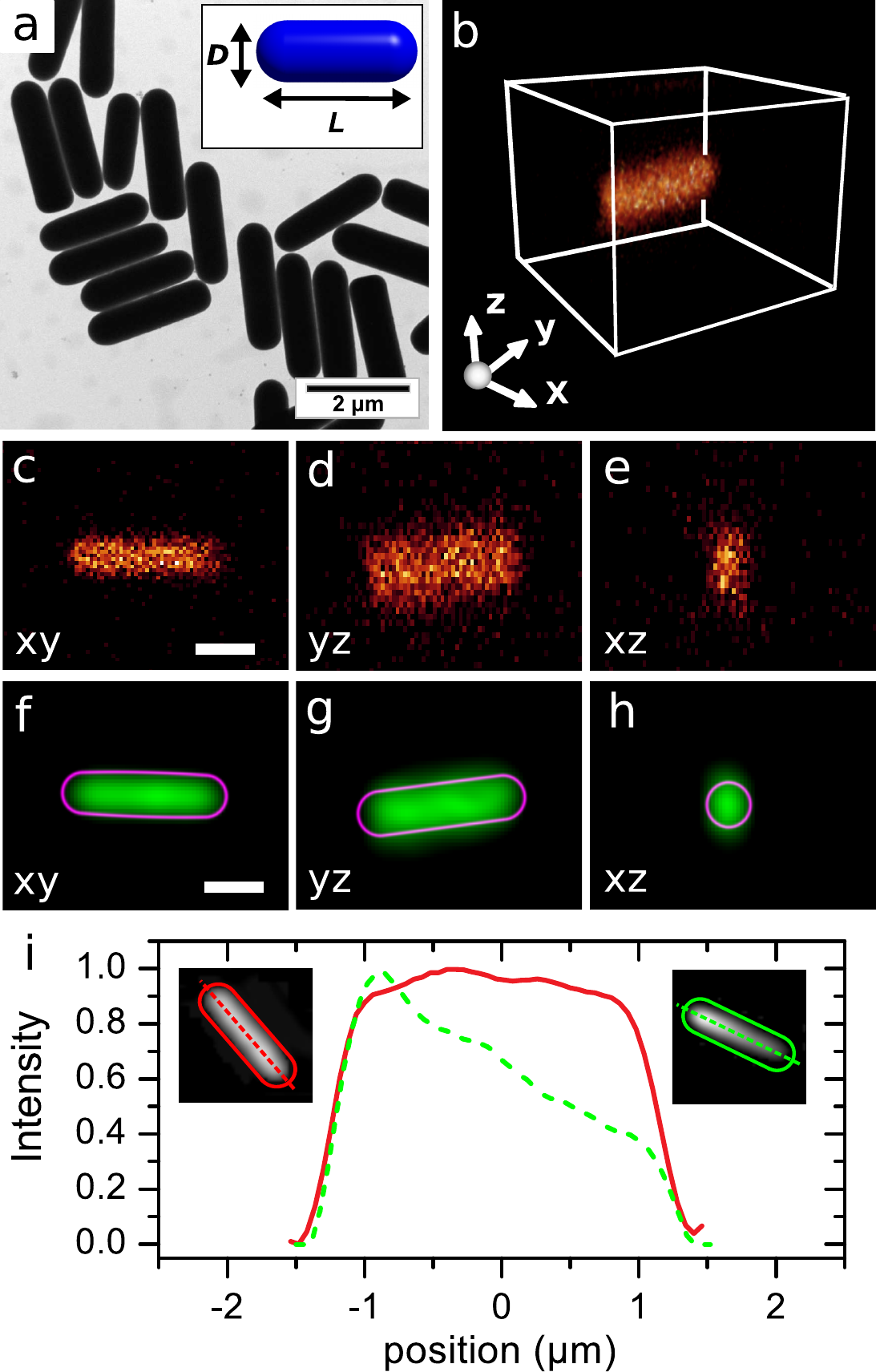}
\caption{ \label{fig:particles}
Particle fitting in 3D. 
(a) Transmission electron microscopy (TEM) micrograph
of fluorescently labelled silica rods with length $l = 2.37$ $\mu$m ($\delta$ = 10\%) and
diameter $d$ = 640 nm ($\delta$ = 7.5\%). 
(b) 3D confocal microscopy image of a single rod suspended in 85wt\%
glycerol in water. 
(c-e) Three orthogonal slices through the 3D confocal image shown in (b). The scale bar is 800 nm. 
(f-h) The particle after rescaling, filtering and
fitting. The magenta outline indicates the final fit from which the position and orientation is computed.  
(i) The (normalized) intensity histograms along the
major axis of two differently dyed particles that are oriented in the $xy$ plane, obtained from
confocal microscopy images. The (red) solid line is from a uniformly dyed
silica rod. The (green) dashed line is from a silica rod with a gradient in dye
distribution.} 
\end{figure}

For the preparation of dense samples of silica rods, two different batches of particles were
used. The first batch consisted of rods with length $l = 2.37$ $\mu$m ($\delta$
= 10\%) and diameter $d$ = 640 nm ($\delta$ = 7.5\%), with $\delta$ the
polydispersity (standard deviation over the mean) \cite{Kuijk2011}. A
transmission electron microscopy (TEM) image of these particles is shown
in Fig.~\ref{fig:particles}a. The particles contained a non-fluorescent core,
a 30 nm fluorescein isothiocyanate (FITC) labelled shell, and a 190 nm
non-fluorescent outer shell. For the second batch of silica rods, with length
$l = 2.6$ $\mu$m (8.5\%) and diameter $d$ = 630 nm (6.3\%), rhodamine
isothiocyanate (RITC) dye was added during synthesis, which resulted in an
intensity gradient of dye molecules along the major axis of the particle
\cite{Kuijk2014a}. The particles were coated with a 175 nm non-fluorescent outer
shell. Particle suspensions were prepared by dispersing the rods in an 
index-matching mixture ($n_D^{21} = 1.45$) of either dimethylsulfoxide (DMSO) and
ultrapure water (Millipore system) or glycerol and ultrapure water. The
particles were first  dispersed in DMSO or glycerol, after which water was
added until the suspension was index-matched by eye. This resulted in mixtures
of 91 wt\% DMSO in water and 85 wt\% glycerol in water. 

Next, sample cells were constructed with standard microscopy slides and No.\
1.0-1.5 glass coverslips (Menzel-Gl\"{a}zer). After the cells were filled with the
suspension, they were sealed with UV-glue (Norland No.\ 68). The suspensions
were imaged  with a confocal microscope (Leica SP2 or Leica SP8) using a
63x/1.4 or 100x/1.4 oil-immersion confocal objective (Leica). We corrected the 3D images
for distortion of the axial (\textit{z}) distances due to the refractive index
mismatch between sample ($n_D^{21}$ = 1.45) and immersion oil ($n_D^{21}$ = 1.51), which resulted in an increase of axial distances of 5\% \cite{Besseling2014}. Fig.~\ref{fig:particles}b shows a 3D confocal microscopy image  
of a single rod suspended in 85 wt\% glycerol in water. In Figs.~\ref{fig:particles}c-e, three orthogonal slices
through this 3D volume are shown. The larger width of the PSF in the axial ($z$)
direction is clearly visible. Notice that the pixel size in \textit{x,y} (50 nm) is
smaller than in \textit{z} (78 nm). Figs.~\ref{fig:particles}f-h show the same rod after
rescaling to cubic pixels, filtering and particle-fitting. In Fig.~\ref{fig:particles}i, we show the intensity histograms of two rods that were oriented parallel
to the \textit{xy} image plane of the confocal microscope. The continuous (red) line shows the intensity histogram of a single uniformly dyed rod and the dashed (green) line that of a
gradient-dyed rod \cite{Kuijk2014a}.
\subsubsection{Freely diffusing silica rods}
For the experimental measurements on a dilute suspension of silica rods we used particles with length $l$ = 3.3 $\mu$m ($\delta$ = 10\%) and diameter $d$ = 550 nm ($\delta$ = 11\%), as measured with TEM. The particles were fluorescently labelled with a 30 nm (FITC) shell. The particles were dispersed in an index matching mixture of 85 wt\% glycerol in water. The density of the solvent mixture was $\rho = 1.222$ g/ml \cite{Segur1951} and the viscosity $\eta = 92$ cP (22$^\circ$C), as measured with an SV10 viscometer (A\&D Company). This mixture not only matches the refractive index of the particles ($n_D^{25}=1.45$), the high viscosity slows down the particle dynamics enough to measure their short-time self-diffusion in 3D. Because the density of this mixture is significantly lower than the density of the particles $\rho = 1.9$ g/ml \cite{Kuijk2014a}, sedimentation cannot be avoided. We assume, however, complete decoupling between translational motion, rotational motion and sedimentation \cite{Homogeneous1979}. A fused quartz capillary (Vitrocom) was filled with a dilute suspension (volume fraction $\phi < 1\%$) of the fluorescent silica rods. The suspension was imaged with a confocal microscope (Leica SP8) equipped with a fast 12 kHz resonant scanner and hybrid detector. Images with 8-bit pixel-depth were acquired using a white light laser with a selected wavelength of 488 nm. A confocal glycerol immersion objective 63x/1.3 (Leica) was used, which is optimized for refractive index $n_D = 1.45$. If we assume a Poisson distribution of the noise, we can easily estimate the signal to noise ratio (SNR) of a single image because of the photon counting mode of the hybrid detector. We use the definition SNR = $\sqrt{n_p}$  with $n_p$ the number of detected photons in the brightest part of the image \cite{Sheppard2006}. 
To avoid hydrodynamic interactions with the wall, particles were imaged 20 $\mu$m deep into the sample. We recorded 800 repeats of 3D image stacks consisting of 512 x 261 x 66 pixels with voxel size 144 x 144 x 331 nm. The time to record a single 3D volume was $\tau = 1.80$ s. During this time, the particles are expected to translate on average $\sqrt{2\, D_t \, \tau} = 110$ nm in each direction and rotate only $\sqrt{4\, D_r \, \tau} = 0.1$ rad.

\subsubsection{AuNRs@SiO$_2$ \& 3D electron tomography}
For the fabrication of a spherical cluster of nanorods, we first synthesized gold nanorods following the method described in Ref.~\citenum{Ye2013}. Next, the gold rods were coated with a layer of mesoporous silica (AuNRs@SiO$_2$) \cite{Gorelikov2008}, which resulted in particles with length $l = 119$ nm and diameter $d = 68$ nm, as measured with TEM. Afterwards, clusters were fabricated via an emulsification process \cite{Peng2013,deNijs2014}. Brightfield TEM tilt series of an 11-particle NR-cluster were acquired by tilting the sample over a range of -65$^\circ$ to 65$^\circ$ and recording images every 2$^\circ$. Images were taken on a Tecnai 20 (FEI) transmission electron microscope, operating at 200 kV with an LaB$_6$ electron source, in bright field mode. Tomographic reconstructions of the images were made with the iMOD software package using the simultaneous iterative reconstruction technique (SIRT) \cite{Kremer1996,Mastronarde1997}. After reconstruction, the data stack was filtered using a low frequency Fourier filter (iMOD) and inverted to ensure light particles on a dark background to enable individual particle identification.

\section{Results}
\subsection{Determination of 3D particle positions \& orientations in dense suspensions}
\begin{figure}[h!]
\includegraphics[width=0.45\textwidth]{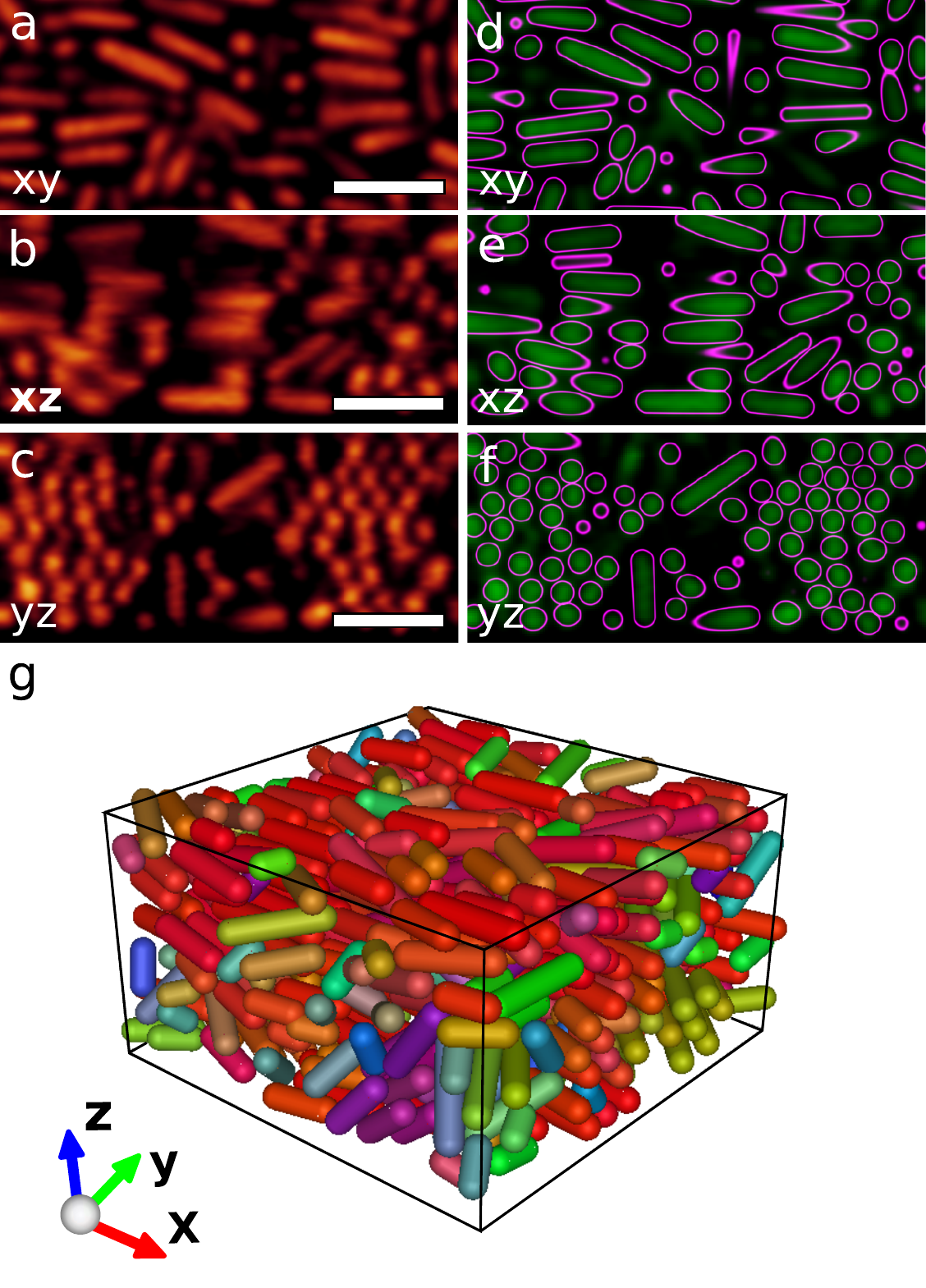}
\caption{Local order in a dense sediment of rods with
length $l$ = 2.37 $\mu$m (10\%), diameter $d$ = 640 nm (7.5 \%) and aspect
ratio $l/d$ = 3.7, dispersed in a glycerol/water mixture. The dimensions of the
image volume were 512 x 201 x 79 pixels with voxel sizes 60 x 60 x 83 nm in $x$,$y$ and $z$.
The time to record the complete stack was 3.37 s. 
(a-c) Close-ups of orthogonal slices through the 3D image, after filtering and (d-f) after particle
identification. The scale bars are 3 $\mu$m. (g) Computer rendered 3D
reconstruction of the sample with the RGB value of the colour indicating the particle orientations.
\label{fig:dense_B31}}
\end{figure}

\begin{figure}
\includegraphics[width=0.45\textwidth]{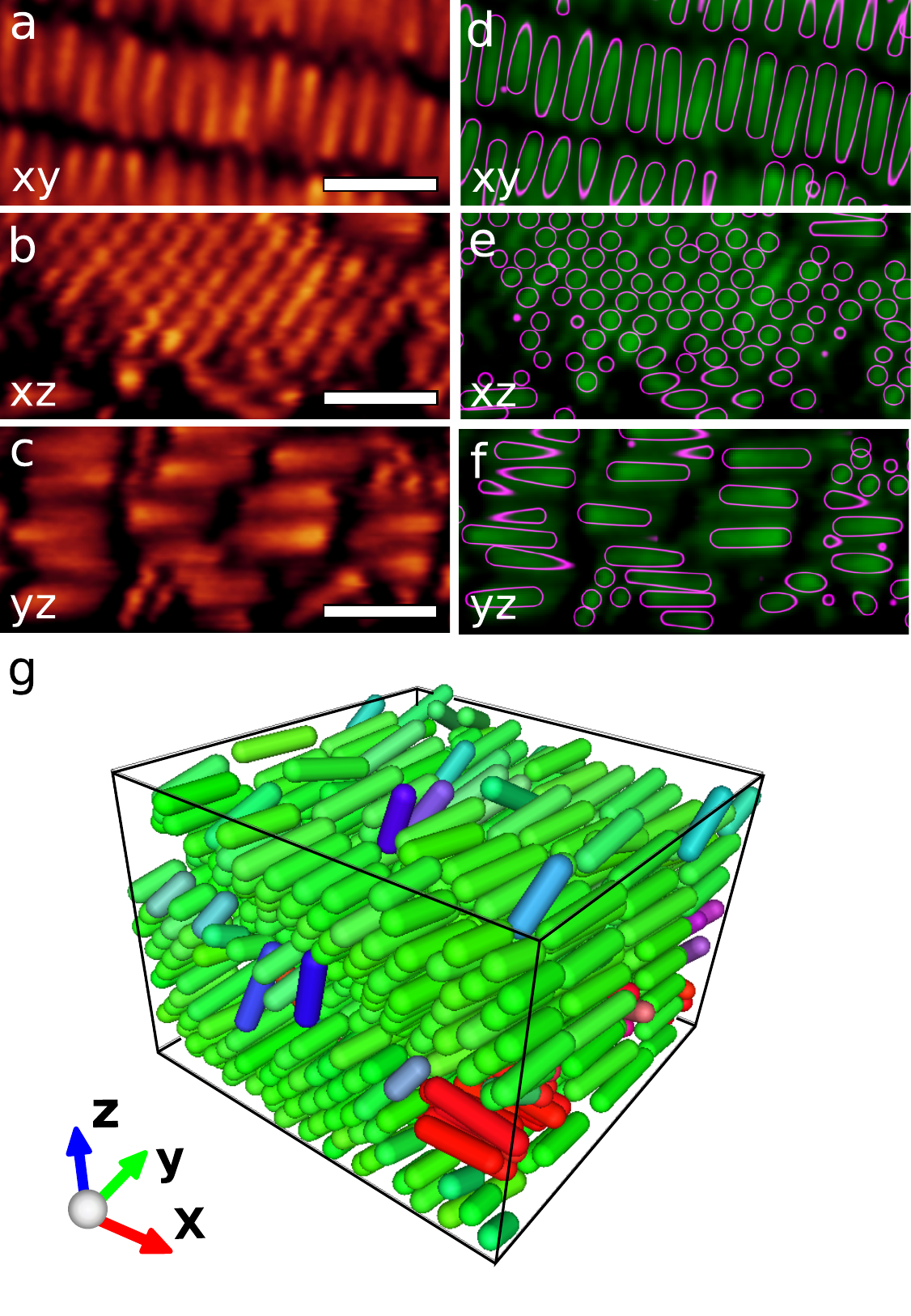}
\caption{Smectic-B phase of rods with length $l$ = 2.6 $\mu$m (8.5 \%), diameter $d$ = 630 nm (6.3 \%) and aspect ratio $l/d$ = 4.1, dispersed in a DMSO/water mixture.  The dimensions of the image volume were 256 x 256 x 151 pixels with voxel size 58 x 58 x 104 nm in $x$,$y$ and $z$. The time to record the image stack was 73.3 s. (a-c) Orthogonal sections after filtering. (a) 12.5 $\mu$m deep in the sample, particles were ordered in smectic-like layers. (d-f) Identified particles are outlined in magenta. All scale bars are 3 $\mu$m. (g)  Computer rendered 3D reconstruction of the sample with the RGB value of the colour indicating the particle orientations.
\label{fig:dense_gradient}}
\end{figure}

To test our 3D particle-fitting algorithm we identified the fluorescent particles in a concentrated suspension of silica rods, as shown in Fig.~\ref{fig:dense_B31}. The particles were uniformly dyed, had a length $l$ = 2.37 $\mu$m (10\%), diameter $d$ = 640 nm (7.5 \%) aspect ratio $l/d$ = 3.7 and were dispersed in a 85 wt\% glycerol in water mixture. Small regions of hexagonally stacked particles existed in the sample (Fig.~\ref{fig:dense_B31}c), however there was no long-ranged order in the sample and particles seemed jammed or arrested in different orientations. Fig.~\ref{fig:dense_B31}a shows that 5.2 $\mu$m deep in the sample, the fluorescent signals of the particles did not overlap significantly in \textit{xy}, despite the high particle concentration. This is due to the 190 nm non-fluorescent outer shell of the particles which was deliberately grown around the particles during synthesis to resolve them individually, even when they were lying side-by-side. However, the orthogonal slices in Figs.~\ref{fig:dense_B31}b-c show that particle signals did overlap in the \textit{z}-direction, even after noise filtering. Nevertheless, by visual inspection of the (magenta) particle outlines in Figs.~\ref{fig:dense_B31}d-f we conclude that the algorithm correctly identified the orientations and positions of the particles, despite the high particle concentration. 
Fig.~\ref{fig:dense_B31}g shows a computer generated reconstruction of the sample, with colours indicating the 3D  orientation of the particles. 
\\ \indent Fig.~\ref{fig:dense_gradient} shows a second example of the performance of our fitting-algorithm in a concentrated suspension. The rods in this sample had length $l$ = 2.6 $\mu$m (8.5 \%), diameter $d$ = 630 nm (6.3 \%) and were dispersed in an index-matching mixture of DMSO/water. After the particles had been left to sediment for several days, they ordered into smectic layers, more or less parallel to the \textit{xy}-plane, as can be seen from Fig.~\ref{fig:dense_gradient}a (12.5 $\mu$m deep in the sample). It can also be seen that the particles had an intensity gradient along their major axis and that there was significant overlap of the fluorescent signals in the \textit{xy}-image (Fig.~\ref{fig:dense_gradient}a). As expected, it was even more difficult to resolve individual particles in the $z$-direction (Figs.~\ref{fig:dense_gradient}b-c), however, it is clear from the hexagonal pattern in Fig.~\ref{fig:dense_gradient}b that the particles formed a smectic-B phase. The magenta outlines in Fig. \ref{fig:dense_gradient}d-f show the result of the particle fitting. By visual inspection of the outlines in the complete image-stack (containing 1699 particles), we conclude that $>$ 98\% of the particles had been correctly identified by the algorithm. In Fig.~\ref{fig:dense_gradient}g we show a 3D reconstruction of a part of the image-stack, which clearly shows 3D orientational order, smectic layering, and transverse (red and blue) particles.   
\subsection{Determination of 3D positions and orientations of gold nanorods}
\begin{figure}[ht!]
\includegraphics[width=0.45\textwidth]{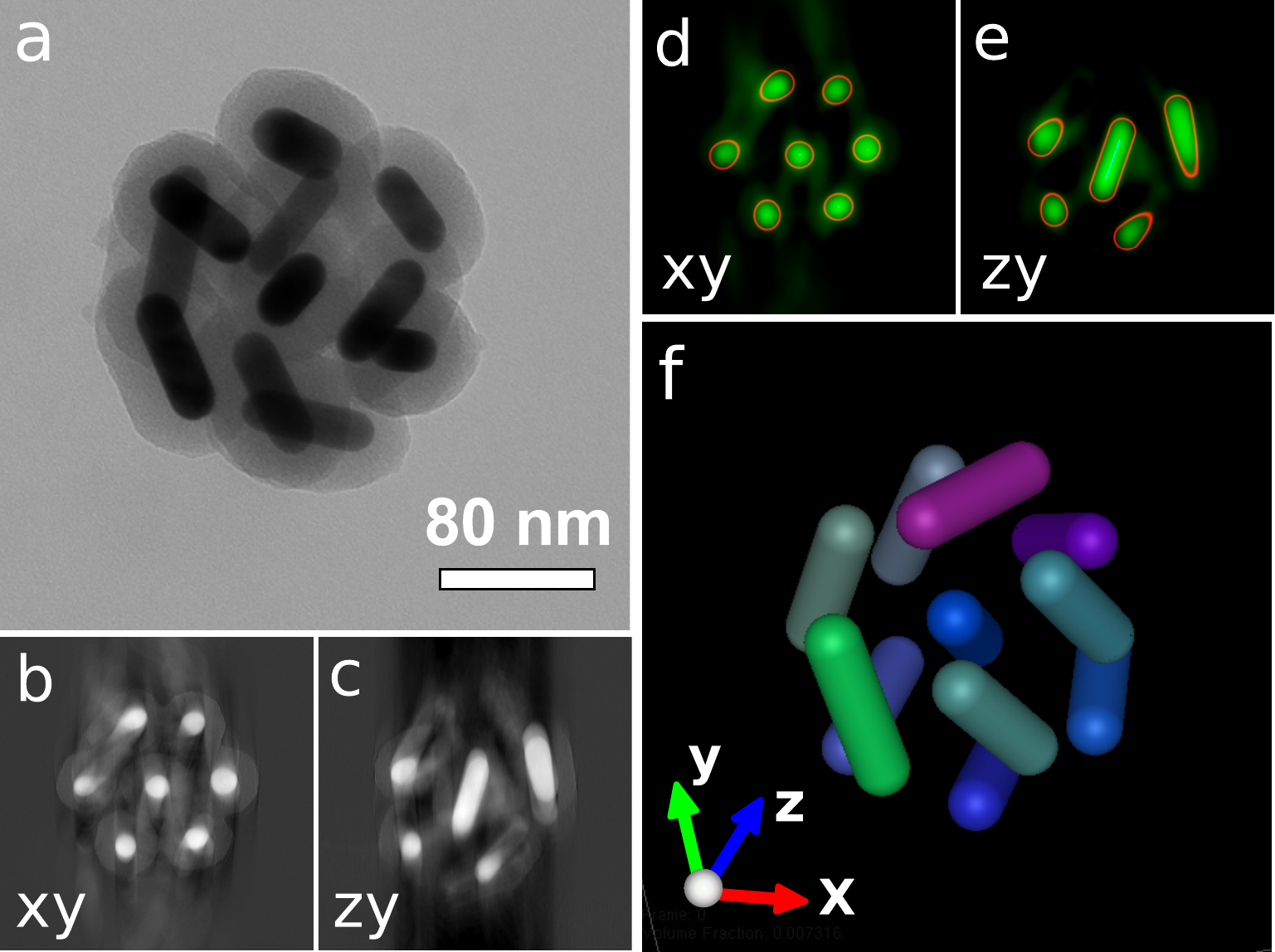}
\caption{
\label{fig:gold_rods}
Identification of the positions and orientations of 11 gold nanorods coated with mesoporous silica (AuNRs@SiO$_2$), confined in a small spherical cluster. 
(a) A single TEM image that was part of the tilt-series used for the tomographic reconstruction. 
(b) An \textit{xy} and (c) \textit{zy}-view of the 3D electron tomogram. The images were inverted for particle identification. 
(d) Corresponding \textit{xy} and (e) \textit{zy}-views of the filtered images with the identified particles outlined in red.
(f) 3D reconstruction of the nanorod cluster. Colours indicate their 3D orientations. 
}
\end{figure}

Although our algorithm was written for analysis of confocal microscopy images, it is also applicable to other 3D  image-stacks of uniaxial symmetric particles. As an example, we show results of the identification of gold nanorods (AuNRs) from a 3D transmission electron microscopy (TEM) tomographic reconstruction in Fig.~\ref{fig:gold_rods}. The TEM micrograph in Fig.~\ref{fig:gold_rods}a shows the gold nanorods (in black), that were coated with a layer of mesoporous silica (dark grey).
Fig.~\ref{fig:gold_rods}b-c show two orthogonal sections through the 3D reconstruction of the cluster. The images were inverted to enable particle identification with our algorithm. Fig.~\ref{fig:gold_rods}d-e show the same orthogonal sections after filtering, with identified particles outlined in red. Finally, Fig.~\ref{fig:gold_rods}f shows the 3D reconstruction, with color-coding of the 3D orientation of the rods. The algorithm had identified all 11 particles and the reconstruction clearly shows that there was some degree of orientational ordering inside the cluster. \\ 

We are aware that a substantial amount of information on the 3D structure of the nanoparticles can be measured directly (and manually) from the 3D tomogram itself. Our image-processing algorithm however can determine unambiguously the 3D positions and orientations of the particles and can therefore be useful for the quantification of (larger) nanoparticle assemblies and should in principle also work on other types of samples, e.g.\ self-assembled clusters of nano-dumbbells \cite{Grzelczak2012}.

\subsection{Testing the accuracy of the algorithm for non-overlapping particle signals}
In this section, we assess the accuracy of our algorithm in more detail. We focus on the fitting accuracy of the algorithm when applied to images containing particle signals that are well separated. Although this situation is much less demanding compared to partially overlapping signals, care has to be taken when fitting this type of data as well. The main reason is that the (fluorescent) diameter of typical rod-like particles
used in our experiments ($d_{fl} \sim 300$ nm) is comparable to the resolution of a typical confocal microscope ($200 - 300$ nm in the lateral and $500 - 700$ nm in the axial direction \cite{Wilson1990,Besseling2014}). Additionally, the PSF itself is anisotropic, which can result in a (strongly) distorted particle shape. Things become progressively worse when there is a refractive-index mismatch between the sample and immersion fluid, which deteriorates the PSF, introduces an intensity fall-off with height and distorts axial distances \cite{Hell1993,Besseling2014}.

We therefore determined the accuracy of our algorithm using two approaches. In the first approach we investigated both the effect of the theoretically approximated PSF and the effect of noise on particle tracking accuracy using computer generated data. The second approach consisted of an experimental measurement of the translational and rotational diffusion of a dilute suspension of silica rods. \\

Details on the construction of the test-images and variation of the theoretically approximated PSF and noise can be found in the Supplementary Information. The results are summarized in Table \ref{tab:errors}, which shows that for our worst-case scenario of a $z$-resolution of 636 nm and signal to noise ratio of 1.7, we obtain for the error in the determination of the main-axis of the rod $\epsilon_r = 0.07$ rad, which corresponds to a small measurement error of $4.1^\circ$, see also Fig.~S1. Additionally, we did not find any significant pixel-bias in either the position or the orientation (see Fig.~S2). For the error in the positional measurement, we found $\epsilon_t/d$ $\sim$ 0.05, which indicates sub-pixel accuracy.

\begin{table} [h]
	\caption{Static measurement error $\epsilon_r$ in the determination of the main axis of the rod, assuming $d$ = 300 nm. The error increases with both $\sigma_z$ and $\sigma_n$. For the worst case scenario of $\sigma_z/d = 0.9$ and $\sigma_n = 0.27$, the value for $\epsilon_r$ remains rather small.}
	\label{tab:errors}
		\begin{tabular}{ c c c c c }
		\multicolumn{2}{c}{z-resolution} & 	\multicolumn{2}{c}{noise levels} & \multicolumn{1}{c}{error} 
		\\\hline 
		\, $\sigma_z/d$ \, & FWHM (nm) & \, $\sigma_n$ \, & SNR & \, $\epsilon_r$ (rad) \, \\
			0.3 & 212  & 0.09 & 13.5 & 0.025 \\
			0.6 & 424  & 0.09 & 11.4 & 0.026 \\
			0.9 & 636  & 0.09 & 11.2 & 0.036 \\
			0.9 & 636  & 0.18 & 3.8 & 0.048 \\ 
			0.9 & 636  & 0.27 & 1.7 & 0.071 \\
		\hline	
		\end{tabular}
	\end{table}

\begin{figure*} [ht!]
\includegraphics[width=1.0\textwidth]{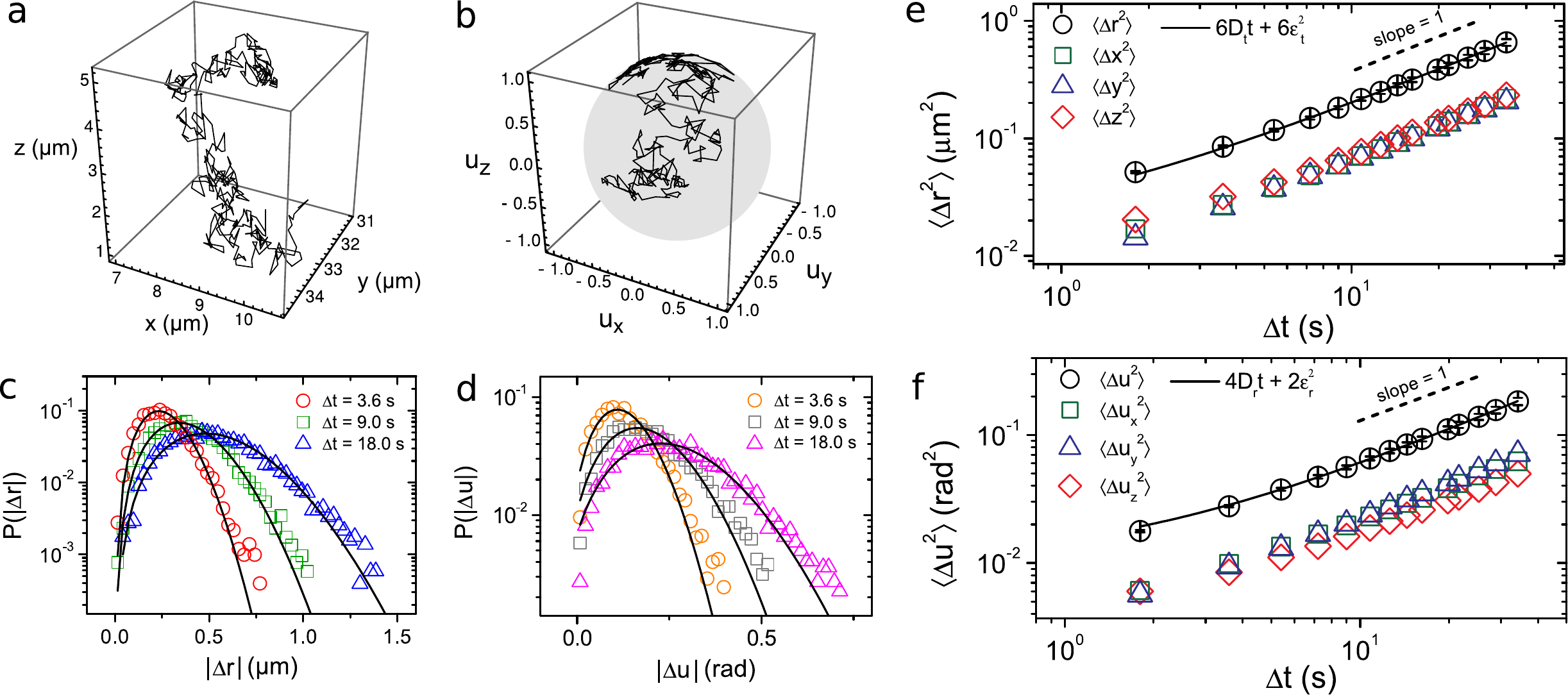}
\caption{
\label{fig:track_exp} 
Experimental measurement on a dilute suspension of sedimenting silica rods with length $l$ = 3.3 $\mu$m and diameter $l$ = 550 nm suspended in a 85wt\% glycerol in water mixture. 
(a) Typical translational and (b) rotational trajectory of a single particle. 
(c) Distribution of the translational displacements $|\Delta {\mathbf r}|$ and (d) rotational displacements $|\Delta {\mathbf{\hat  u}}|$ for three different time-steps $\Delta t$. The displacements are an average over 8 particles and the black lines are fits. 
(e) The average mean squared displacement (MSD). The estimate for the static error ($\epsilon_t = 45, 46$ and $59$ nm in $x$, $y$ and $z$ respectively) confirms sub-pixel accuracy. 
(f) Mean squared angular displacement (MSAD). The static error in the determination of the unit vector $\epsilon_r = 0.07$ rad corresponds to an angular uncertainty of $4^\circ$. Both the fitted translational and rotational diffusion coefficients are in good agreement with the analytical predictions from Ref.~\citenum{Tirado1984}.}
\end{figure*}

Finally, we measured the diffusive motion in a dilute suspension of fluorescent silica rods experimentally, which provides a real-life test of the accuracy of our algorithm. The rods that were used had length $l$ = 3.3 $\mu$m ($\delta$ = 10\%), diameter $d$ = 550 nm ($\delta$ = 11\%) and aspect ratio $l$/$d$ = 6.0. From the number of photons in the brightest part of the image we estimated the signal to noise ratio to be SNR $\approx 3$, which is  in the range stated in Table \ref{tab:errors}. The tracking results, averaged over 8 particles, are shown in Fig.~\ref{fig:track_exp}. A typical translational trajectory of 12 min is shown in Fig.~\ref{fig:track_exp}a. From a fit to the average linear displacements (of all 8 particles) in the z-direction, we estimate the sedimentation speed to be $v_{sed} = 0.331\,\pm\,$0.005 $\mu$m/min. This value is slightly higher but comparable to the value of $v_{sed} = 0.28$ $\mu$m/min that we obtained from equation \eqref{eq:sed}. For further analysis we subtracted the average linear displacements from the trajectories. Fig.~\ref{fig:track_exp}b shows a rotational trajectory of 12 min for a single particle. In Fig.~\ref{fig:track_exp}c we show the probability distribution of the norm of the displacement $|\Delta {\mathbf r}|$, for three different time-steps $\Delta$t.   In Fig.~\ref{fig:track_exp}d we show the same distribution for the norm of the displacements of the unit vector $|\Delta {\mathbf{\hat  u}}|$. The solid black lines in Figs.~\ref{fig:track_exp}c,d are fits proportional to $|\alpha|^2 \exp(-|\alpha|^2)$ with $\alpha = {\Delta\mathbf r},{\Delta\mathbf{\hat  u}}$ respectively. 

To extract the translational diffusion coefficient, we calculated the rotationally averaged mean squared displacement $\langle \Delta \mathbf{r}^2 \rangle$, as can be seen in Fig.~\ref{fig:track_exp}e. For $\Delta t > 10$ s we found that $\langle \Delta \mathbf{r}^2 \rangle \sim t^{0.97}$ indicating diffusive behaviour. The statistical error in the individual measurement points is smaller than the symbol size. Fitting the data with  equation \eqref{eq:msd}, we obtain the short-time rotationally averaged translation diffusion coefficient $D_t = (3.06\,\pm\,0.01)\,\times\,10^{-3}$ $\mu$m$^2$/s and static error $\epsilon_t = 45, 46$ and $59$ nm in the $x$, $y$ and $z$-direction respectively, which confirms that we can locate the particles with sub-pixel accuracy. The value for $D_t$ is in strong  agreement with the theoretical value obtained from equation \eqref{eq:dtrans} which is $D_t$ = 3.2 $\times\,10^{-3}$ $\mu$m$^2$/s. Finally, we calculated the mean squared angular displacement $\langle \Delta \mathbf{\hat {u}}^2 \rangle$, as shown in Fig.~\ref{fig:track_exp}f. This time we obtained  $\langle \Delta \mathbf{\hat {u}}^2 \rangle \sim t^{0.92}$ and for the short-time rotational diffusion coefficient $D_r$ = (1.32$\,\pm\,0.02)\,\times\,10^{-3}$ rad$^2$/s. This is in good agreement with the theoretical value $D_r$ = 1.5 $\times\,10^{-3}$ rad$^2$/s, obtained from equation \eqref{eq:Dr}. For the corresponding rotational relaxation time we found $\tau_r = 1/(2 \, D_r)$ = 3.8 $\times\,10^2$ s, which confirms that we measured in the short-time diffusion regime. From the fit we also obtained $\epsilon_r$ = 0.07 rad, which corresponds to a small angular uncertainty of $\sim$ 4$^\circ$.

\section{Discussion}
In this paper we demonstrated a new image-processing algorithm that is
capable of extracting the positions and orientations of fluorescent rod-like
particles in both dilute and concentrated suspensions. Although the algorithm was written for three
dimensions, most steps are straightforward to modify for two dimensions \cite{Besseling}. Mohraz and Solomon
\cite{Mohraz2005} were the first, as far as we know, to describe an algorithm that can detect the 
position and orientation of ellipsoidal particles in 3D confocal microscopy
images and this work follows a similar approach. The algorithm of Mohraz and Solomon groups clusters of pixels together to form backbones but 
does not use additional fitting steps, which we found necessary to
correctly identify particles when there is significant overlap of particle
signals. The difference in particle geometry (ellipsoids versus rods) combined
with the small (fluorescent) particle diameter in our study might be the reason why we find
that using only a maximum threshold and cluster analysis is not sufficient to
identify rods in concentrated suspensions, even when the rods have a large ($>
150$ nm) non-fluorescent shell and a considerable electric double layer ($\sim
100$ nm) \cite{Kuijk2011,Kuijk2012a}. The rod-like particles used in this study have a 
repulsive interaction potential and therefore form dense smectic-like phases, which we now can identify on  the single-particle level in the bulk. The algorithm also enables the study of glassy phases of anisotropic particles in three dimensions, which is promising since all current real-space
glass-transition studies of anisotropic particles so far are either 2D
\cite{Yunker2011a,Zheng2011,Mishra2013} or tracer-host \cite{Edmond2012}.
Finally, we would like to mention that the algorithm is also applicable to study
the dynamics of (concentrated) `active colloids' (e.g.\ self-propelled particles
and bacteria), a field that is rapidly emerging \cite{Ball2013}. 
\\  \indent Since the typical fluorescent diameter of the rod-like particles is around 300
nm, deconvolution of the image-stacks before particle fitting can be useful
when particles are difficult to resolve individually. The necessary higher (Nyquist) 
sampling rate, however, is not always practical or even not possible for faster
moving particles. Additionally, deconvolutions are sensitive to small changes in the rod-spread-function (RSP), introduced by e.g.\ polydispersity. A clear improvement of the algorithm, therefore, is to fit the particles
with the RSP, analogous to the fitting of the sphere-spread-function (SSF) reported by Jenkins \textit{et al.}
\cite{Jenkins2008}, which is work currently ongoing. With this type of extension of the current algorithm it should also be possible to accurately measure in-situ particle polydispersity, which is known to have a large effect on e.g.\ the liquid-crystalline phase behaviour \cite{Bates1998}. 
\\ \indent By measurement of freely diffusing rods, we
acquired additional information on the accuracy of our algorithm. Although the
motion is analysed in the lab-frame and therefore translational and rotational
motion should be coupled \cite{Han2006}, we did not observe such behaviour
since the friction anisotropy in our 3D measurement is small
$D_\parallel$/$D_\perp$ = 1.3 and because we averaged over an ensemble of
particles and over many initial orientations. We found that the error in
locating the rods ($\epsilon_t = 45, 46$ and $59$ nm in the $x$, $y$ and $z$-direction respectively) confirms sub-pixel accuracy and agrees
roughly with the criterion for spherical particles that $\epsilon_t \sim M/N$,
with $M$ the pixel-size and $N$ the diameter of the particle in pixels
\cite{Crocker1996}. The value for the (short-time) rotational diffusion
coefficient $D_r$ = (1.32$\,\pm\,0.02)\,\times\,10^{-3}$ rad$^2$/s, is one order of magnitude larger than
previously accessible with 3D confocal microscopy \cite{Mukhija2007} which is,
however,  due to the equipment rather than the image-processing. The error in
the determination of the orientation of the rod ($\epsilon_r$ = 0.07 rad) is in
the range of the values obtained via simulated test-images, shown in Table
\ref{tab:errors}. The rule-of-thumb that $\epsilon_r \sim 1/P_a$ with $P_a$ the
half-length of the rods in pixels \cite{Mukhija2007} seems to hold quite well
in our case, since $1/P_a$ = 0.08 in our measurements. 

\section{Conclusion}
We developed an algorithm that extracts the positions and orientations of rod-like particles from 3D confocal microscopy images. 
The algorithm is tailored to a system of
fluorescently labelled silica rods and can identify these particles even in the bulk of 3D
concentrated phases where the fluorescent signals of the particles overlap considerably. This allowed us to determine the 3D positions and orientations of particles 
in a concentrated disordered phase and in a liquid-crystalline smectic-B phase.
The algorithm also works on electron tomography
reconstructions of gold nanorods, which enables the 3D quantification of
(large) nano-particle assemblies. It is also expected to work on other uniaxial particles such as ellipsoids or 
dumbbells. 
We determined the accuracy of the algorithm
for varying $z$-resolution and noise levels from generated 3D test-images. Despite the (anisotropic) distortion of the theoretically approximated point spread function (PSF) and
the low signal to noise ratio (SNR), the error in the determination of the
orientation of the particles remained small. These results confirmed that we can
accurately track rod-like particles with (fluorescent) diameters down to 300 nm. With our
algorithm and a fast confocal microscope we determined the translational and
rotational motion of a dilute suspension of sedimenting silica rods. We
demonstrated that the measured  diffusive motion was in good agreement with
theory (neglecting sedimentation) and that we can track the particles with sub-pixel resolution.
\\ \indent This novel algorithm therefore allows for studies of structure and dynamics on the particle level of dense  liquid-crystalline phase behaviour (such as nematic, smectic and crystalline
phases), but also allows for studies of the glass transition of anisotropic
rod-like particles in three dimensions. Of course, the algorithm will also be applicable to dilute 
suspensions or in cases where rod-like particles are used as tracers, such as in biophysical or micro-rheology 
studies.

\subsection{Acknowledgement} 
The authors thank E. Weeks and A. Gantapara for useful
discussion, H. Bakker, C. Kennedy and B. Liu for useful feedback
on the algorithm and H. Meeldijk for help with the electron microscopy data. 
This research was carried out partially (THB) under project
number M62.7.08SDMP25 in the framework of the Industrial Partnership Program on
Size Dependent Material Properties of the Materials innovation institute (M2i)
and the Foundation of Fundamental Research on Matter (FOM), which is part of
the Netherlands Organisation for Scientific Research (NWO).  Part of
the research leading to these results has received funding from the European Research
Council under the European Union’s Seventh Framework Programme (FP/2007-2013)/
ERC Grant Agreement no. [291667].

\subsection{Author Contributions} 
AvB initiated the research. MD, AI and AvB supervised the research. MH and THB contributed equally to this work. MH and THB wrote and tested the particle fitting algorithm and analysed the data. AK performed synthesis of the fluorescent silica rods. TSD performed synthesis of the gold nanorods, fabrication of the nanorod-clusters and acquired the tilt-series of TEM images. BdN performed the electron tomography reconstruction. THB and AK performed the confocal microscopy experiments. THB, MH and AvB co-wrote the manuscript.

\bibliography{./library_rod_fitting_3D.bib}

\begin{thebibliography}{65}%
\makeatletter
\providecommand \@ifxundefined [1]{%
 \@ifx{#1\undefined}
}%
\providecommand \@ifnum [1]{%
 \ifnum #1\expandafter \@firstoftwo
 \else \expandafter \@secondoftwo
 \fi
}%
\providecommand \@ifx [1]{%
 \ifx #1\expandafter \@firstoftwo
 \else \expandafter \@secondoftwo
 \fi
}%
\providecommand \natexlab [1]{#1}%
\providecommand \enquote  [1]{``#1''}%
\providecommand \bibnamefont  [1]{#1}%
\providecommand \bibfnamefont [1]{#1}%
\providecommand \citenamefont [1]{#1}%
\providecommand \href@noop [0]{\@secondoftwo}%
\providecommand \href [0]{\begingroup \@sanitize@url \@href}%
\providecommand \@href[1]{\@@startlink{#1}\@@href}%
\providecommand \@@href[1]{\endgroup#1\@@endlink}%
\providecommand \@sanitize@url [0]{\catcode `\\12\catcode `\$12\catcode
  `\&12\catcode `\#12\catcode `\^12\catcode `\_12\catcode `\%12\relax}%
\providecommand \@@startlink[1]{}%
\providecommand \@@endlink[0]{}%
\providecommand \url  [0]{\begingroup\@sanitize@url \@url }%
\providecommand \@url [1]{\endgroup\@href {#1}{\urlprefix }}%
\providecommand \urlprefix  [0]{URL }%
\providecommand \Eprint [0]{\href }%
\providecommand \doibase [0]{http://dx.doi.org/}%
\providecommand \selectlanguage [0]{\@gobble}%
\providecommand \bibinfo  [0]{\@secondoftwo}%
\providecommand \bibfield  [0]{\@secondoftwo}%
\providecommand \translation [1]{[#1]}%
\providecommand \BibitemOpen [0]{}%
\providecommand \bibitemStop [0]{}%
\providecommand \bibitemNoStop [0]{.\EOS\space}%
\providecommand \EOS [0]{\spacefactor3000\relax}%
\providecommand \BibitemShut  [1]{\csname bibitem#1\endcsname}%
\let\auto@bib@innerbib\@empty
\bibitem [{\citenamefont {Croll}(2002)}]{Croll2002}%
  \BibitemOpen
  \bibfield  {author} {\bibinfo {author} {\bibfnamefont {S.}~\bibnamefont
  {Croll}},\ }\href
  {http://www.sciencedirect.com/science/article/pii/S0300944001002612}
  {\bibfield  {journal} {\bibinfo  {journal} {Prog. Org. Coatings}\ }\textbf
  {\bibinfo {volume} {44}},\ \bibinfo {pages} {131} (\bibinfo {year}
  {2002})}\BibitemShut {NoStop}%
\bibitem [{\citenamefont {Letchford}\ and\ \citenamefont
  {Burt}(2007)}]{Letchford2007}%
  \BibitemOpen
  \bibfield  {author} {\bibinfo {author} {\bibfnamefont {K.}~\bibnamefont
  {Letchford}}\ and\ \bibinfo {author} {\bibfnamefont {H.}~\bibnamefont
  {Burt}},\ }\href {\doibase 10.1016/j.ejpb.2006.11.009} {\bibfield  {journal}
  {\bibinfo  {journal} {Eur. J. Pharm. Biopharm}\ }\textbf {\bibinfo {volume}
  {65}},\ \bibinfo {pages} {259} (\bibinfo {year} {2007})}\BibitemShut
  {NoStop}%
\bibitem [{\citenamefont {Sozer}\ and\ \citenamefont
  {Kokini}(2009)}]{Sozer2009}%
  \BibitemOpen
  \bibfield  {author} {\bibinfo {author} {\bibfnamefont {N.}~\bibnamefont
  {Sozer}}\ and\ \bibinfo {author} {\bibfnamefont {J.~L.}\ \bibnamefont
  {Kokini}},\ }\href {\doibase 10.1016/j.tibtech.2008.10.010} {\bibfield
  {journal} {\bibinfo  {journal} {Trends Biotechnol.}\ }\textbf {\bibinfo
  {volume} {27}},\ \bibinfo {pages} {82} (\bibinfo {year} {2009})}\BibitemShut
  {NoStop}%
\bibitem [{\citenamefont {Comiskey}\ \emph {et~al.}(1998)\citenamefont
  {Comiskey}, \citenamefont {Albert}, \citenamefont {Yoshizawa},\ and\
  \citenamefont {Jacobsen}}]{Sucher1998}%
  \BibitemOpen
  \bibfield  {author} {\bibinfo {author} {\bibfnamefont {B.}~\bibnamefont
  {Comiskey}}, \bibinfo {author} {\bibfnamefont {J.}~\bibnamefont {Albert}},
  \bibinfo {author} {\bibfnamefont {H.}~\bibnamefont {Yoshizawa}}, \ and\
  \bibinfo {author} {\bibfnamefont {J.}~\bibnamefont {Jacobsen}},\ }\href@noop
  {} {\bibfield  {journal} {\bibinfo  {journal} {Nature}\ }\textbf {\bibinfo
  {volume} {394}},\ \bibinfo {pages} {253} (\bibinfo {year}
  {1998})}\BibitemShut {NoStop}%
\bibitem [{\citenamefont {van Blaaderen}\ and\ \citenamefont
  {Wiltzius}(1995)}]{VanBlaaderen1995}%
  \BibitemOpen
  \bibfield  {author} {\bibinfo {author} {\bibfnamefont {A.}~\bibnamefont {van
  Blaaderen}}\ and\ \bibinfo {author} {\bibfnamefont {P.}~\bibnamefont
  {Wiltzius}},\ }\href {\doibase 10.1126/science.270.5239.1177} {\bibfield
  {journal} {\bibinfo  {journal} {Science}\ }\textbf {\bibinfo {volume}
  {270}},\ \bibinfo {pages} {1177} (\bibinfo {year} {1995})}\BibitemShut
  {NoStop}%
\bibitem [{\citenamefont {Weeks}\ \emph {et~al.}(2000)\citenamefont {Weeks},
  \citenamefont {Crocker}, \citenamefont {Levitt}, \citenamefont {Schofield},\
  and\ \citenamefont {Weitz}}]{Weeks2000}%
  \BibitemOpen
  \bibfield  {author} {\bibinfo {author} {\bibfnamefont {E.~R.}\ \bibnamefont
  {Weeks}}, \bibinfo {author} {\bibfnamefont {J.~C.}\ \bibnamefont {Crocker}},
  \bibinfo {author} {\bibfnamefont {A.}~\bibnamefont {Levitt}}, \bibinfo
  {author} {\bibfnamefont {A.}~\bibnamefont {Schofield}}, \ and\ \bibinfo
  {author} {\bibfnamefont {D.~A.}\ \bibnamefont {Weitz}},\ }\href {\doibase
  10.1126/science.287.5453.627} {\bibfield  {journal} {\bibinfo  {journal}
  {Science}\ }\textbf {\bibinfo {volume} {287}},\ \bibinfo {pages} {627}
  (\bibinfo {year} {2000})}\BibitemShut {NoStop}%
\bibitem [{\citenamefont {Dassanayake}\ \emph {et~al.}(2000)\citenamefont
  {Dassanayake}, \citenamefont {Fraden},\ and\ \citenamefont {van
  Blaaderen}}]{Dassanayake2000}%
  \BibitemOpen
  \bibfield  {author} {\bibinfo {author} {\bibfnamefont {U.}~\bibnamefont
  {Dassanayake}}, \bibinfo {author} {\bibfnamefont {S.}~\bibnamefont {Fraden}},
  \ and\ \bibinfo {author} {\bibfnamefont {A.}~\bibnamefont {van Blaaderen}},\
  }\href {\doibase 10.1063/1.480933} {\bibfield  {journal} {\bibinfo  {journal}
  {J. Chem. Phys.}\ }\textbf {\bibinfo {volume} {112}},\ \bibinfo {pages}
  {3851} (\bibinfo {year} {2000})}\BibitemShut {NoStop}%
\bibitem [{\citenamefont {Gasser}\ \emph {et~al.}(2001)\citenamefont {Gasser},
  \citenamefont {Weeks}, \citenamefont {Schofield}, \citenamefont {Pusey},\
  and\ \citenamefont {Weitz}}]{Gasser2001}%
  \BibitemOpen
  \bibfield  {author} {\bibinfo {author} {\bibfnamefont {U.}~\bibnamefont
  {Gasser}}, \bibinfo {author} {\bibfnamefont {E.~R.}\ \bibnamefont {Weeks}},
  \bibinfo {author} {\bibfnamefont {A.}~\bibnamefont {Schofield}}, \bibinfo
  {author} {\bibfnamefont {P.~N.}\ \bibnamefont {Pusey}}, \ and\ \bibinfo
  {author} {\bibfnamefont {D.~A.}\ \bibnamefont {Weitz}},\ }\href {\doibase
  10.1126/science.1058457} {\bibfield  {journal} {\bibinfo  {journal}
  {Science}\ }\textbf {\bibinfo {volume} {292}},\ \bibinfo {pages} {258}
  (\bibinfo {year} {2001})}\BibitemShut {NoStop}%
\bibitem [{\citenamefont {Dinsmore}\ \emph {et~al.}(2001)\citenamefont
  {Dinsmore}, \citenamefont {Weeks}, \citenamefont {Prasad}, \citenamefont
  {Levitt},\ and\ \citenamefont {Weitz}}]{Dinsmore2001}%
  \BibitemOpen
  \bibfield  {author} {\bibinfo {author} {\bibfnamefont {A.~D.}\ \bibnamefont
  {Dinsmore}}, \bibinfo {author} {\bibfnamefont {E.~R.}\ \bibnamefont {Weeks}},
  \bibinfo {author} {\bibfnamefont {V.}~\bibnamefont {Prasad}}, \bibinfo
  {author} {\bibfnamefont {A.~C.}\ \bibnamefont {Levitt}}, \ and\ \bibinfo
  {author} {\bibfnamefont {D.~A.}\ \bibnamefont {Weitz}},\ }\href
  {http://www.ncbi.nlm.nih.gov/pubmed/18360451} {\bibfield  {journal} {\bibinfo
   {journal} {Appl. Opt.}\ }\textbf {\bibinfo {volume} {40}},\ \bibinfo {pages}
  {4152} (\bibinfo {year} {2001})}\BibitemShut {NoStop}%
\bibitem [{\citenamefont {Cohen}\ \emph {et~al.}(2006)\citenamefont {Cohen},
  \citenamefont {Davidovitch}, \citenamefont {Schofield}, \citenamefont
  {Brenner},\ and\ \citenamefont {Weitz}}]{Cohen2006}%
  \BibitemOpen
  \bibfield  {author} {\bibinfo {author} {\bibfnamefont {I.}~\bibnamefont
  {Cohen}}, \bibinfo {author} {\bibfnamefont {B.}~\bibnamefont {Davidovitch}},
  \bibinfo {author} {\bibfnamefont {A.~B.}\ \bibnamefont {Schofield}}, \bibinfo
  {author} {\bibfnamefont {M.~P.}\ \bibnamefont {Brenner}}, \ and\ \bibinfo
  {author} {\bibfnamefont {D.~A.}\ \bibnamefont {Weitz}},\ }\href
  {http://link.aps.org/doi/10.1103/PhysRevLett.97.215502} {\bibfield  {journal}
  {\bibinfo  {journal} {Phys. Rev. Lett.}\ }\textbf {\bibinfo {volume} {97}},\
  \bibinfo {pages} {215502} (\bibinfo {year} {2006})}\BibitemShut {NoStop}%
\bibitem [{\citenamefont {Yethiraj}\ and\ \citenamefont {van
  Blaaderen}(2003)}]{Yethiraj2003}%
  \BibitemOpen
  \bibfield  {author} {\bibinfo {author} {\bibfnamefont {A.}~\bibnamefont
  {Yethiraj}}\ and\ \bibinfo {author} {\bibfnamefont {A.}~\bibnamefont {van
  Blaaderen}},\ }\href {\doibase 10.1038/nature01328} {\bibfield  {journal}
  {\bibinfo  {journal} {Nature}\ }\textbf {\bibinfo {volume} {421}},\ \bibinfo
  {pages} {513} (\bibinfo {year} {2003})}\BibitemShut {NoStop}%
\bibitem [{\citenamefont {Besseling}\ \emph {et~al.}(2012)\citenamefont
  {Besseling}, \citenamefont {Hermes}, \citenamefont {Fortini}, \citenamefont
  {Dijkstra}, \citenamefont {Imhof},\ and\ \citenamefont {van
  Blaaderen}}]{Besseling2012a}%
  \BibitemOpen
  \bibfield  {author} {\bibinfo {author} {\bibfnamefont {T.~H.}\ \bibnamefont
  {Besseling}}, \bibinfo {author} {\bibfnamefont {M.}~\bibnamefont {Hermes}},
  \bibinfo {author} {\bibfnamefont {A.}~\bibnamefont {Fortini}}, \bibinfo
  {author} {\bibfnamefont {M.}~\bibnamefont {Dijkstra}}, \bibinfo {author}
  {\bibfnamefont {A.}~\bibnamefont {Imhof}}, \ and\ \bibinfo {author}
  {\bibfnamefont {A.}~\bibnamefont {van Blaaderen}},\ }\href {\doibase
  10.1039/c2sm07156h} {\bibfield  {journal} {\bibinfo  {journal} {Soft Matter}\
  }\textbf {\bibinfo {volume} {8}},\ \bibinfo {pages} {6931} (\bibinfo {year}
  {2012})}\BibitemShut {NoStop}%
\bibitem [{\citenamefont {Crocker}\ and\ \citenamefont
  {Grier}(1996)}]{Crocker1996}%
  \BibitemOpen
  \bibfield  {author} {\bibinfo {author} {\bibfnamefont {J.~C.}\ \bibnamefont
  {Crocker}}\ and\ \bibinfo {author} {\bibfnamefont {D.~G.}\ \bibnamefont
  {Grier}},\ }\href@noop {} {\bibfield  {journal} {\bibinfo  {journal} {J.
  Colloid Interface Sci.}\ }\textbf {\bibinfo {volume} {179}},\ \bibinfo
  {pages} {298} (\bibinfo {year} {1996})}\BibitemShut {NoStop}%
\bibitem [{\citenamefont {Lu}\ \emph {et~al.}(2007)\citenamefont {Lu},
  \citenamefont {Sims}, \citenamefont {Oki}, \citenamefont {Macarthur},\ and\
  \citenamefont {Weitz}}]{Lu2007}%
  \BibitemOpen
  \bibfield  {author} {\bibinfo {author} {\bibfnamefont {P.~J.}\ \bibnamefont
  {Lu}}, \bibinfo {author} {\bibfnamefont {P.~A.}\ \bibnamefont {Sims}},
  \bibinfo {author} {\bibfnamefont {H.}~\bibnamefont {Oki}}, \bibinfo {author}
  {\bibfnamefont {J.~B.}\ \bibnamefont {Macarthur}}, \ and\ \bibinfo {author}
  {\bibfnamefont {D.~A.}\ \bibnamefont {Weitz}},\ }\href
  {http://www.ncbi.nlm.nih.gov/pubmed/19547205} {\bibfield  {journal} {\bibinfo
   {journal} {Opt. Express}\ }\textbf {\bibinfo {volume} {15}},\ \bibinfo
  {pages} {8702} (\bibinfo {year} {2007})}\BibitemShut {NoStop}%
\bibitem [{\citenamefont {Jenkins}\ and\ \citenamefont
  {Egelhaaf}(2008)}]{Jenkins2008}%
  \BibitemOpen
  \bibfield  {author} {\bibinfo {author} {\bibfnamefont {M.~C.}\ \bibnamefont
  {Jenkins}}\ and\ \bibinfo {author} {\bibfnamefont {S.~U.}\ \bibnamefont
  {Egelhaaf}},\ }\href
  {http://www.sciencedirect.com/science/article/pii/S0001868607001200}
  {\bibfield  {journal} {\bibinfo  {journal} {Adv. Colloid Interface Sci.}\
  }\textbf {\bibinfo {volume} {136}},\ \bibinfo {pages} {65} (\bibinfo {year}
  {2008})}\BibitemShut {NoStop}%
\bibitem [{\citenamefont {Gao}\ and\ \citenamefont {Kilfoil}(2009)}]{Gao2009}%
  \BibitemOpen
  \bibfield  {author} {\bibinfo {author} {\bibfnamefont {Y.}~\bibnamefont
  {Gao}}\ and\ \bibinfo {author} {\bibfnamefont {M.}~\bibnamefont {Kilfoil}},\
  }\href {\doibase 10.1007/s11265-008-0250-2} {\bibfield  {journal} {\bibinfo
  {journal} {Opt. Express}\ }\textbf {\bibinfo {volume} {17}},\ \bibinfo
  {pages} {4685} (\bibinfo {year} {2009})}\BibitemShut {NoStop}%
\bibitem [{\citenamefont {Besseling}\ \emph {et~al.}(2009)\citenamefont
  {Besseling}, \citenamefont {Isa}, \citenamefont {Weeks},\ and\ \citenamefont
  {Poon}}]{Besseling2009a}%
  \BibitemOpen
  \bibfield  {author} {\bibinfo {author} {\bibfnamefont {R.}~\bibnamefont
  {Besseling}}, \bibinfo {author} {\bibfnamefont {L.}~\bibnamefont {Isa}},
  \bibinfo {author} {\bibfnamefont {E.~R.}\ \bibnamefont {Weeks}}, \ and\
  \bibinfo {author} {\bibfnamefont {W.~C.}\ \bibnamefont {Poon}},\ }\href@noop
  {} {\bibfield  {journal} {\bibinfo  {journal} {Adv. Colloid Interface Sci.}\
  }\textbf {\bibinfo {volume} {146}},\ \bibinfo {pages} {1} (\bibinfo {year}
  {2009})}\BibitemShut {NoStop}%
\bibitem [{\citenamefont {Vissers}\ \emph {et~al.}(2011)\citenamefont
  {Vissers}, \citenamefont {Imhof}, \citenamefont {Carrique}, \citenamefont
  {Delgado},\ and\ \citenamefont {van Blaaderen}}]{Vissers2011}%
  \BibitemOpen
  \bibfield  {author} {\bibinfo {author} {\bibfnamefont {T.}~\bibnamefont
  {Vissers}}, \bibinfo {author} {\bibfnamefont {A.}~\bibnamefont {Imhof}},
  \bibinfo {author} {\bibfnamefont {F.}~\bibnamefont {Carrique}}, \bibinfo
  {author} {\bibfnamefont {A.~V.}\ \bibnamefont {Delgado}}, \ and\ \bibinfo
  {author} {\bibfnamefont {A.}~\bibnamefont {van Blaaderen}},\ }\href {\doibase
  10.1016/j.jcis.2011.04.113} {\bibfield  {journal} {\bibinfo  {journal} {J.
  Colloid Interface Sci.}\ }\textbf {\bibinfo {volume} {361}},\ \bibinfo
  {pages} {443} (\bibinfo {year} {2011})}\BibitemShut {NoStop}%
\bibitem [{\citenamefont {Kurita}\ \emph {et~al.}(2012)\citenamefont {Kurita},
  \citenamefont {Ruffner},\ and\ \citenamefont {Weeks}}]{Kurita2012a}%
  \BibitemOpen
  \bibfield  {author} {\bibinfo {author} {\bibfnamefont {R.}~\bibnamefont
  {Kurita}}, \bibinfo {author} {\bibfnamefont {D.~B.}\ \bibnamefont {Ruffner}},
  \ and\ \bibinfo {author} {\bibfnamefont {E.~R.}\ \bibnamefont {Weeks}},\
  }\href {\doibase 10.1038/ncomms2114} {\bibfield  {journal} {\bibinfo
  {journal} {Nat. Commun.}\ }\textbf {\bibinfo {volume} {3}},\ \bibinfo {pages}
  {1127} (\bibinfo {year} {2012})}\BibitemShut {NoStop}%
\bibitem [{\citenamefont {Leocmach}\ and\ \citenamefont
  {Tanaka}(2013)}]{Leocmach2013}%
  \BibitemOpen
  \bibfield  {author} {\bibinfo {author} {\bibfnamefont {M.}~\bibnamefont
  {Leocmach}}\ and\ \bibinfo {author} {\bibfnamefont {H.}~\bibnamefont
  {Tanaka}},\ }\href {\doibase 10.1039/c2sm27107a} {\bibfield  {journal}
  {\bibinfo  {journal} {Soft Matter}\ }\textbf {\bibinfo {volume} {9}},\
  \bibinfo {pages} {1447} (\bibinfo {year} {2013})}\BibitemShut {NoStop}%
\bibitem [{\citenamefont {Kuijk}\ \emph {et~al.}(2011)\citenamefont {Kuijk},
  \citenamefont {van Blaaderen},\ and\ \citenamefont {Imhof}}]{Kuijk2011}%
  \BibitemOpen
  \bibfield  {author} {\bibinfo {author} {\bibfnamefont {A.}~\bibnamefont
  {Kuijk}}, \bibinfo {author} {\bibfnamefont {A.}~\bibnamefont {van
  Blaaderen}}, \ and\ \bibinfo {author} {\bibfnamefont {A.}~\bibnamefont
  {Imhof}},\ }\href {\doibase 10.1021/ja109524h} {\bibfield  {journal}
  {\bibinfo  {journal} {J. Am. Chem. Soc.}\ }\textbf {\bibinfo {volume}
  {133}},\ \bibinfo {pages} {2346} (\bibinfo {year} {2011})}\BibitemShut
  {NoStop}%
\bibitem [{\citenamefont {Kuijk}\ \emph {et~al.}(2012)\citenamefont {Kuijk},
  \citenamefont {Byelov}, \citenamefont {Petukhov}, \citenamefont {van
  Blaaderen},\ and\ \citenamefont {Imhof}}]{Kuijk2012a}%
  \BibitemOpen
  \bibfield  {author} {\bibinfo {author} {\bibfnamefont {A.}~\bibnamefont
  {Kuijk}}, \bibinfo {author} {\bibfnamefont {D.~V.}\ \bibnamefont {Byelov}},
  \bibinfo {author} {\bibfnamefont {A.~V.}\ \bibnamefont {Petukhov}}, \bibinfo
  {author} {\bibfnamefont {A.}~\bibnamefont {van Blaaderen}}, \ and\ \bibinfo
  {author} {\bibfnamefont {A.}~\bibnamefont {Imhof}},\ }\href {\doibase
  10.1039/c2fd20084h} {\bibfield  {journal} {\bibinfo  {journal} {Faraday
  Discuss.}\ }\textbf {\bibinfo {volume} {159}},\ \bibinfo {pages} {181}
  (\bibinfo {year} {2012})}\BibitemShut {NoStop}%
\bibitem [{\citenamefont {Mohraz}\ and\ \citenamefont
  {Solomon}(2005)}]{Mohraz2005}%
  \BibitemOpen
  \bibfield  {author} {\bibinfo {author} {\bibfnamefont {A.}~\bibnamefont
  {Mohraz}}\ and\ \bibinfo {author} {\bibfnamefont {M.~J.}\ \bibnamefont
  {Solomon}},\ }\href {\doibase 10.1021/la046908a} {\bibfield  {journal}
  {\bibinfo  {journal} {Langmuir}\ }\textbf {\bibinfo {volume} {21}},\ \bibinfo
  {pages} {5298} (\bibinfo {year} {2005})}\BibitemShut {NoStop}%
\bibitem [{\citenamefont {Rossi}\ \emph {et~al.}(2011)\citenamefont {Rossi},
  \citenamefont {Sacanna}, \citenamefont {Irvine}, \citenamefont {Chaikin},
  \citenamefont {Pine},\ and\ \citenamefont {Philipse}}]{Rossi2011}%
  \BibitemOpen
  \bibfield  {author} {\bibinfo {author} {\bibfnamefont {L.}~\bibnamefont
  {Rossi}}, \bibinfo {author} {\bibfnamefont {S.}~\bibnamefont {Sacanna}},
  \bibinfo {author} {\bibfnamefont {W.~T.~M.}\ \bibnamefont {Irvine}}, \bibinfo
  {author} {\bibfnamefont {P.~M.}\ \bibnamefont {Chaikin}}, \bibinfo {author}
  {\bibfnamefont {D.~J.}\ \bibnamefont {Pine}}, \ and\ \bibinfo {author}
  {\bibfnamefont {A.~P.}\ \bibnamefont {Philipse}},\ }\href {\doibase
  10.1039/c0sm01246g} {\bibfield  {journal} {\bibinfo  {journal} {Soft Matter}\
  }\textbf {\bibinfo {volume} {7}},\ \bibinfo {pages} {4139} (\bibinfo {year}
  {2011})}\BibitemShut {NoStop}%
\bibitem [{\citenamefont {Elsesser}\ \emph {et~al.}(2011)\citenamefont
  {Elsesser}, \citenamefont {Hollingsworth}, \citenamefont {Edmond},\ and\
  \citenamefont {Pine}}]{Elsesser2011}%
  \BibitemOpen
  \bibfield  {author} {\bibinfo {author} {\bibfnamefont {M.~T.}\ \bibnamefont
  {Elsesser}}, \bibinfo {author} {\bibfnamefont {A.~D.}\ \bibnamefont
  {Hollingsworth}}, \bibinfo {author} {\bibfnamefont {K.~V.}\ \bibnamefont
  {Edmond}}, \ and\ \bibinfo {author} {\bibfnamefont {D.~J.}\ \bibnamefont
  {Pine}},\ }\href {\doibase 10.1021/la1034905} {\bibfield  {journal} {\bibinfo
   {journal} {Langmuir}\ }\textbf {\bibinfo {volume} {27}},\ \bibinfo {pages}
  {917} (\bibinfo {year} {2011})}\BibitemShut {NoStop}%
\bibitem [{\citenamefont {Peng}\ \emph {et~al.}(2012)\citenamefont {Peng},
  \citenamefont {Vutukuri}, \citenamefont {van Blaaderen},\ and\ \citenamefont
  {Imhof}}]{Peng2012}%
  \BibitemOpen
  \bibfield  {author} {\bibinfo {author} {\bibfnamefont {B.}~\bibnamefont
  {Peng}}, \bibinfo {author} {\bibfnamefont {H.~R.}\ \bibnamefont {Vutukuri}},
  \bibinfo {author} {\bibfnamefont {A.}~\bibnamefont {van Blaaderen}}, \ and\
  \bibinfo {author} {\bibfnamefont {A.}~\bibnamefont {Imhof}},\ }\href
  {\doibase 10.1039/c2jm35229j} {\bibfield  {journal} {\bibinfo  {journal} {J.
  Mater. Chem.}\ }\textbf {\bibinfo {volume} {22}},\ \bibinfo {pages} {21893}
  (\bibinfo {year} {2012})}\BibitemShut {NoStop}%
\bibitem [{\citenamefont {Anthony}\ \emph {et~al.}(2006)\citenamefont
  {Anthony}, \citenamefont {Hong}, \citenamefont {Kim},\ and\ \citenamefont
  {Granick}}]{Anthony2006}%
  \BibitemOpen
  \bibfield  {author} {\bibinfo {author} {\bibfnamefont {S.~M.}\ \bibnamefont
  {Anthony}}, \bibinfo {author} {\bibfnamefont {L.}~\bibnamefont {Hong}},
  \bibinfo {author} {\bibfnamefont {M.}~\bibnamefont {Kim}}, \ and\ \bibinfo
  {author} {\bibfnamefont {S.}~\bibnamefont {Granick}},\ }\href {\doibase
  10.1021/la062094h} {\bibfield  {journal} {\bibinfo  {journal} {Langmuir}\
  }\textbf {\bibinfo {volume} {22}},\ \bibinfo {pages} {9812} (\bibinfo {year}
  {2006})}\BibitemShut {NoStop}%
\bibitem [{\citenamefont {Han}\ \emph {et~al.}(2006)\citenamefont {Han},
  \citenamefont {Alsayed}, \citenamefont {Nobili}, \citenamefont {Zhang},
  \citenamefont {Lubensky},\ and\ \citenamefont {Yodh}}]{Han2006}%
  \BibitemOpen
  \bibfield  {author} {\bibinfo {author} {\bibfnamefont {Y.}~\bibnamefont
  {Han}}, \bibinfo {author} {\bibfnamefont {A.~M.}\ \bibnamefont {Alsayed}},
  \bibinfo {author} {\bibfnamefont {M.}~\bibnamefont {Nobili}}, \bibinfo
  {author} {\bibfnamefont {J.}~\bibnamefont {Zhang}}, \bibinfo {author}
  {\bibfnamefont {T.~C.}\ \bibnamefont {Lubensky}}, \ and\ \bibinfo {author}
  {\bibfnamefont {A.~G.}\ \bibnamefont {Yodh}},\ }\href {\doibase
  10.1126/science.1130146} {\bibfield  {journal} {\bibinfo  {journal}
  {Science}\ }\textbf {\bibinfo {volume} {314}},\ \bibinfo {pages} {626}
  (\bibinfo {year} {2006})}\BibitemShut {NoStop}%
\bibitem [{\citenamefont {Hunter}\ \emph {et~al.}(2011)\citenamefont {Hunter},
  \citenamefont {Edmond}, \citenamefont {Elsesser},\ and\ \citenamefont
  {Weeks}}]{Hunter2011}%
  \BibitemOpen
  \bibfield  {author} {\bibinfo {author} {\bibfnamefont {G.~L.}\ \bibnamefont
  {Hunter}}, \bibinfo {author} {\bibfnamefont {K.~V.}\ \bibnamefont {Edmond}},
  \bibinfo {author} {\bibfnamefont {M.~T.}\ \bibnamefont {Elsesser}}, \ and\
  \bibinfo {author} {\bibfnamefont {E.~R.}\ \bibnamefont {Weeks}},\ }\href
  {http://www.ncbi.nlm.nih.gov/pubmed/21935082} {\bibfield  {journal} {\bibinfo
   {journal} {Opt. Express}\ }\textbf {\bibinfo {volume} {19}},\ \bibinfo
  {pages} {17189} (\bibinfo {year} {2011})}\BibitemShut {NoStop}%
\bibitem [{\citenamefont {Zhao}\ \emph {et~al.}(2011)\citenamefont {Zhao},
  \citenamefont {Bruinsma},\ and\ \citenamefont {Mason}}]{Zhao2011}%
  \BibitemOpen
  \bibfield  {author} {\bibinfo {author} {\bibfnamefont {K.}~\bibnamefont
  {Zhao}}, \bibinfo {author} {\bibfnamefont {R.}~\bibnamefont {Bruinsma}}, \
  and\ \bibinfo {author} {\bibfnamefont {T.~G.}\ \bibnamefont {Mason}},\ }\href
  {\doibase 10.1073/pnas.1014942108} {\bibfield  {journal} {\bibinfo  {journal}
  {Proc. Natl. Acad. Sci. U.S.A.}\ }\textbf {\bibinfo {volume} {108}},\
  \bibinfo {pages} {2684} (\bibinfo {year} {2011})}\BibitemShut {NoStop}%
\bibitem [{\citenamefont {Chakrabarty}\ \emph {et~al.}(2013)\citenamefont
  {Chakrabarty}, \citenamefont {Wang}, \citenamefont {Fan}, \citenamefont
  {Sun},\ and\ \citenamefont {Wei}}]{Chakrabarty2013}%
  \BibitemOpen
  \bibfield  {author} {\bibinfo {author} {\bibfnamefont {A.}~\bibnamefont
  {Chakrabarty}}, \bibinfo {author} {\bibfnamefont {F.}~\bibnamefont {Wang}},
  \bibinfo {author} {\bibfnamefont {C.-Z.}\ \bibnamefont {Fan}}, \bibinfo
  {author} {\bibfnamefont {K.}~\bibnamefont {Sun}}, \ and\ \bibinfo {author}
  {\bibfnamefont {Q.-H.}\ \bibnamefont {Wei}},\ }\href {\doibase
  10.1021/la403427y} {\bibfield  {journal} {\bibinfo  {journal} {Langmuir}\
  }\textbf {\bibinfo {volume} {29}},\ \bibinfo {pages} {14396} (\bibinfo {year}
  {2013})}\BibitemShut {NoStop}%
\bibitem [{\citenamefont {Mukhija}\ and\ \citenamefont
  {Solomon}(2007)}]{Mukhija2007}%
  \BibitemOpen
  \bibfield  {author} {\bibinfo {author} {\bibfnamefont {D.}~\bibnamefont
  {Mukhija}}\ and\ \bibinfo {author} {\bibfnamefont {M.~J.}\ \bibnamefont
  {Solomon}},\ }\href {\doibase 10.1016/j.jcis.2007.05.055} {\bibfield
  {journal} {\bibinfo  {journal} {J. Colloid Interface Sci.}\ }\textbf
  {\bibinfo {volume} {314}},\ \bibinfo {pages} {98} (\bibinfo {year}
  {2007})}\BibitemShut {NoStop}%
\bibitem [{\citenamefont {Mukhija}\ and\ \citenamefont
  {Solomon}(2010)}]{Mukhija2011}%
  \BibitemOpen
  \bibfield  {author} {\bibinfo {author} {\bibfnamefont {D.}~\bibnamefont
  {Mukhija}}\ and\ \bibinfo {author} {\bibfnamefont {M.~J.}\ \bibnamefont
  {Solomon}},\ }\href {\doibase 10.1039/c0sm00493f} {\bibfield  {journal}
  {\bibinfo  {journal} {Soft Matter}\ }\textbf {\bibinfo {volume} {7}},\
  \bibinfo {pages} {540} (\bibinfo {year} {2010})}\BibitemShut {NoStop}%
\bibitem [{\citenamefont {Shah}\ \emph {et~al.}(2012)\citenamefont {Shah},
  \citenamefont {Kang}, \citenamefont {Kohlstedt}, \citenamefont {Ahn},
  \citenamefont {Glotzer}, \citenamefont {Monroe},\ and\ \citenamefont
  {Solomon}}]{Shah2012}%
  \BibitemOpen
  \bibfield  {author} {\bibinfo {author} {\bibfnamefont {A.~A.}\ \bibnamefont
  {Shah}}, \bibinfo {author} {\bibfnamefont {H.}~\bibnamefont {Kang}}, \bibinfo
  {author} {\bibfnamefont {K.~L.}\ \bibnamefont {Kohlstedt}}, \bibinfo {author}
  {\bibfnamefont {K.~H.}\ \bibnamefont {Ahn}}, \bibinfo {author} {\bibfnamefont
  {S.~C.}\ \bibnamefont {Glotzer}}, \bibinfo {author} {\bibfnamefont {C.~W.}\
  \bibnamefont {Monroe}}, \ and\ \bibinfo {author} {\bibfnamefont {M.~J.}\
  \bibnamefont {Solomon}},\ }\href {\doibase 10.1002/smll.201102265} {\bibfield
   {journal} {\bibinfo  {journal} {Small}\ }\textbf {\bibinfo {volume} {8}},\
  \bibinfo {pages} {1551} (\bibinfo {year} {2012})}\BibitemShut {NoStop}%
\bibitem [{\citenamefont {Kuijk}\ \emph {et~al.}(2014)\citenamefont {Kuijk},
  \citenamefont {Imhof}, \citenamefont {Verkuijlen}, \citenamefont {Besseling},
  \citenamefont {van Eck},\ and\ \citenamefont {van Blaaderen}}]{Kuijk2014a}%
  \BibitemOpen
  \bibfield  {author} {\bibinfo {author} {\bibfnamefont {A.}~\bibnamefont
  {Kuijk}}, \bibinfo {author} {\bibfnamefont {A.}~\bibnamefont {Imhof}},
  \bibinfo {author} {\bibfnamefont {M.~H.~W.}\ \bibnamefont {Verkuijlen}},
  \bibinfo {author} {\bibfnamefont {T.~H.}\ \bibnamefont {Besseling}}, \bibinfo
  {author} {\bibfnamefont {E.~R.~H.}\ \bibnamefont {van Eck}}, \ and\ \bibinfo
  {author} {\bibfnamefont {A.}~\bibnamefont {van Blaaderen}},\ }\href {\doibase
  10.1002/ppsc.201300329} {\bibfield  {journal} {\bibinfo  {journal} {Part.
  Part. Syst. Char.}\ } (\bibinfo {year} {2014}),\
  10.1002/ppsc.201300329}\BibitemShut {NoStop}%
\bibitem [{\citenamefont {Cannell}\ \emph {et~al.}(2006)\citenamefont
  {Cannell}, \citenamefont {Mcmorland},\ and\ \citenamefont
  {Soeller}}]{Cannell2006}%
  \BibitemOpen
  \bibfield  {author} {\bibinfo {author} {\bibfnamefont {M.~B.}\ \bibnamefont
  {Cannell}}, \bibinfo {author} {\bibfnamefont {A.}~\bibnamefont {Mcmorland}},
  \ and\ \bibinfo {author} {\bibfnamefont {C.}~\bibnamefont {Soeller}},\ }in\
  \href@noop {} {\emph {\bibinfo {booktitle} {Handbook Of Biological Confocal
  Microscopy}}},\ \bibinfo {editor} {edited by\ \bibinfo {editor}
  {\bibfnamefont {J.~B.}\ \bibnamefont {Pawley}}}\ (\bibinfo  {publisher}
  {Springer Science+Business Media},\ \bibinfo {address} {New York},\ \bibinfo
  {year} {2006})\ \bibinfo {edition} {3rd}\ ed.,\ Chap.~\bibinfo {chapter}
  {25}, pp.\ \bibinfo {pages} {488--500}\BibitemShut {NoStop}%
\bibitem [{\citenamefont {Besag}\ \emph {et~al.}(1991)\citenamefont {Besag},
  \citenamefont {York},\ and\ \citenamefont {Molli\'{e}}}]{Besag1991}%
  \BibitemOpen
  \bibfield  {author} {\bibinfo {author} {\bibfnamefont {J.}~\bibnamefont
  {Besag}}, \bibinfo {author} {\bibfnamefont {J.}~\bibnamefont {York}}, \ and\
  \bibinfo {author} {\bibfnamefont {A.}~\bibnamefont {Molli\'{e}}},\ }\href
  {http://link.springer.com/article/10.1007/BF00116466} {\bibfield  {journal}
  {\bibinfo  {journal} {Ann. Inst. Statist. Math.}\ }\textbf {\bibinfo {volume}
  {43}},\ \bibinfo {pages} {1} (\bibinfo {year} {1991})}\BibitemShut {NoStop}%
\bibitem [{\citenamefont {Al-Awadhi}\ \emph {et~al.}(2004)\citenamefont
  {Al-Awadhi}, \citenamefont {Jennison},\ and\ \citenamefont
  {Hurn}}]{Al-Awadhi2004}%
  \BibitemOpen
  \bibfield  {author} {\bibinfo {author} {\bibfnamefont {F.}~\bibnamefont
  {Al-Awadhi}}, \bibinfo {author} {\bibfnamefont {C.}~\bibnamefont {Jennison}},
  \ and\ \bibinfo {author} {\bibfnamefont {M.}~\bibnamefont {Hurn}},\ }\href
  {\doibase 10.1046/j.0035-9254.2003.05177.x} {\bibfield  {journal} {\bibinfo
  {journal} {Appl. Statist.}\ }\textbf {\bibinfo {volume} {53}},\ \bibinfo
  {pages} {31} (\bibinfo {year} {2004})}\BibitemShut {NoStop}%
\bibitem [{\citenamefont {Art}(2006)}]{Art2006}%
  \BibitemOpen
  \bibfield  {author} {\bibinfo {author} {\bibfnamefont {J.}~\bibnamefont
  {Art}},\ }in\ \href@noop {} {\emph {\bibinfo {booktitle} {Handbook of
  Biological Confocal Microscopy}}},\ \bibinfo {editor} {edited by\ \bibinfo
  {editor} {\bibfnamefont {J.~B.}\ \bibnamefont {Pawley}}}\ (\bibinfo
  {publisher} {Springer Science+Business Media},\ \bibinfo {address} {New
  York},\ \bibinfo {year} {2006})\ \bibinfo {edition} {3rd}\ ed.,\
  Chap.~\bibinfo {chapter} {12}, pp.\ \bibinfo {pages} {251--264}\BibitemShut
  {NoStop}%
\bibitem [{\citenamefont {Shakarji}(1998)}]{Shakarji1998}%
  \BibitemOpen
  \bibfield  {author} {\bibinfo {author} {\bibfnamefont {C.}~\bibnamefont
  {Shakarji}},\ }\href {\doibase 10.6028/jres.103.043} {\bibfield  {journal}
  {\bibinfo  {journal} {J. Res. Natl. Ins.t Stand. Technol.}\ }\textbf
  {\bibinfo {volume} {103}},\ \bibinfo {pages} {633} (\bibinfo {year}
  {1998})}\BibitemShut {NoStop}%
\bibitem [{\citenamefont {Savin}\ and\ \citenamefont
  {Doyle}(2005)}]{Savin2005}%
  \BibitemOpen
  \bibfield  {author} {\bibinfo {author} {\bibfnamefont {T.}~\bibnamefont
  {Savin}}\ and\ \bibinfo {author} {\bibfnamefont {P.~S.}\ \bibnamefont
  {Doyle}},\ }\href {\doibase 10.1529/biophysj.104.042457} {\bibfield
  {journal} {\bibinfo  {journal} {Biophys. J.}\ }\textbf {\bibinfo {volume}
  {88}},\ \bibinfo {pages} {623} (\bibinfo {year} {2005})}\BibitemShut
  {NoStop}%
\bibitem [{\citenamefont {Dhont}(1996)}]{Dhont1996}%
  \BibitemOpen
  \bibfield  {author} {\bibinfo {author} {\bibfnamefont {J.}~\bibnamefont
  {Dhont}},\ }\href@noop {} {\emph {\bibinfo {title} {{An Introduction to
  Dynamics of Colloids}}}}\ (\bibinfo  {publisher} {Elsevier},\ \bibinfo
  {address} {Amsterdam},\ \bibinfo {year} {1996})\BibitemShut {NoStop}%
\bibitem [{\citenamefont {Cheong}\ and\ \citenamefont
  {Grier}(2010)}]{Cheong2010}%
  \BibitemOpen
  \bibfield  {author} {\bibinfo {author} {\bibfnamefont {F.~C.}\ \bibnamefont
  {Cheong}}\ and\ \bibinfo {author} {\bibfnamefont {D.~G.}\ \bibnamefont
  {Grier}},\ }\href
  {http://www.opticsinfobase.org/oe/fulltext.cfm?uri=oe-18-7-6555\&id=196647}
  {\bibfield  {journal} {\bibinfo  {journal} {Opt. Express}\ }\textbf {\bibinfo
  {volume} {18}},\ \bibinfo {pages} {6555} (\bibinfo {year}
  {2010})}\BibitemShut {NoStop}%
\bibitem [{\citenamefont {Brenner}(1979)}]{Homogeneous1979}%
  \BibitemOpen
  \bibfield  {author} {\bibinfo {author} {\bibfnamefont {H.}~\bibnamefont
  {Brenner}},\ }\href@noop {} {\bibfield  {journal} {\bibinfo  {journal} {J.
  Colloid Interface Sci.}\ }\textbf {\bibinfo {volume} {71}},\ \bibinfo {pages}
  {189} (\bibinfo {year} {1979})}\BibitemShut {NoStop}%
\bibitem [{\citenamefont {Svedberg}\ and\ \citenamefont
  {Pedersen}(1940)}]{Svedberg1940}%
  \BibitemOpen
  \bibfield  {author} {\bibinfo {author} {\bibfnamefont {T.}~\bibnamefont
  {Svedberg}}\ and\ \bibinfo {author} {\bibfnamefont {K.}~\bibnamefont
  {Pedersen}},\ }\href@noop {} {\emph {\bibinfo {title} {{The
  Ultracentrifuge}}}}\ (\bibinfo  {publisher} {Oxford Univ. Press},\ \bibinfo
  {address} {London},\ \bibinfo {year} {1940})\BibitemShut {NoStop}%
\bibitem [{\citenamefont {Tirado}\ \emph {et~al.}(1984)\citenamefont {Tirado},
  \citenamefont {Martínez},\ and\ \citenamefont {de~la Torre}}]{Tirado1984}%
  \BibitemOpen
  \bibfield  {author} {\bibinfo {author} {\bibfnamefont {M.~M.}\ \bibnamefont
  {Tirado}}, \bibinfo {author} {\bibfnamefont {C.~L.}\ \bibnamefont
  {Martínez}}, \ and\ \bibinfo {author} {\bibfnamefont {J.~G.}\ \bibnamefont
  {de~la Torre}},\ }\href {\doibase 10.1063/1.447827} {\bibfield  {journal}
  {\bibinfo  {journal} {J. Chem. Phys.}\ }\textbf {\bibinfo {volume} {81}},\
  \bibinfo {pages} {2047} (\bibinfo {year} {1984})}\BibitemShut {NoStop}%
\bibitem [{\citenamefont {Besseling}\ \emph {et~al.}(2014)\citenamefont
  {Besseling}, \citenamefont {Jose},\ and\ \citenamefont {van
  Blaaderen}}]{Besseling2014}%
  \BibitemOpen
  \bibfield  {author} {\bibinfo {author} {\bibfnamefont {T.~H.}\ \bibnamefont
  {Besseling}}, \bibinfo {author} {\bibfnamefont {J.}~\bibnamefont {Jose}}, \
  and\ \bibinfo {author} {\bibfnamefont {A.}~\bibnamefont {van Blaaderen}},\
  }\href@noop {} {\bibfield  {journal} {\bibinfo  {journal}
  {arXiv:1404.3952v1}\ } (\bibinfo {year} {2014})}\BibitemShut {NoStop}%
\bibitem [{\citenamefont {Segur}\ and\ \citenamefont
  {Oberstar}(1951)}]{Segur1951}%
  \BibitemOpen
  \bibfield  {author} {\bibinfo {author} {\bibfnamefont {J.~B.}\ \bibnamefont
  {Segur}}\ and\ \bibinfo {author} {\bibfnamefont {H.~E.}\ \bibnamefont
  {Oberstar}},\ }\href@noop {} {\bibfield  {journal} {\bibinfo  {journal} {Ind.
  Eng. Chem}\ }\textbf {\bibinfo {volume} {43}},\ \bibinfo {pages} {2117 }
  (\bibinfo {year} {1951})}\BibitemShut {NoStop}%
\bibitem [{\citenamefont {Sheppard}\ \emph {et~al.}(2006)\citenamefont
  {Sheppard}, \citenamefont {Gan}, \citenamefont {Gu},\ and\ \citenamefont
  {Roy}}]{Sheppard2006}%
  \BibitemOpen
  \bibfield  {author} {\bibinfo {author} {\bibfnamefont {C.~J.~R.}\
  \bibnamefont {Sheppard}}, \bibinfo {author} {\bibfnamefont {X.}~\bibnamefont
  {Gan}}, \bibinfo {author} {\bibfnamefont {M.}~\bibnamefont {Gu}}, \ and\
  \bibinfo {author} {\bibfnamefont {M.}~\bibnamefont {Roy}},\ }in\ \href@noop
  {} {\emph {\bibinfo {booktitle} {Handbook of Biological Confocal
  Microscopy}}},\ \bibinfo {editor} {edited by\ \bibinfo {editor}
  {\bibfnamefont {J.~B.}\ \bibnamefont {Pawley}}}\ (\bibinfo  {publisher}
  {Springer Science+Business Media},\ \bibinfo {address} {New York},\ \bibinfo
  {year} {2006})\ \bibinfo {edition} {3rd}\ ed.,\ Chap.~\bibinfo {chapter}
  {22}, pp.\ \bibinfo {pages} {442--452}\BibitemShut {NoStop}%
\bibitem [{\citenamefont {Ye}\ \emph {et~al.}(2013)\citenamefont {Ye},
  \citenamefont {Zheng}, \citenamefont {Chen}, \citenamefont {Gao},\ and\
  \citenamefont {Murray}}]{Ye2013}%
  \BibitemOpen
  \bibfield  {author} {\bibinfo {author} {\bibfnamefont {X.}~\bibnamefont
  {Ye}}, \bibinfo {author} {\bibfnamefont {C.}~\bibnamefont {Zheng}}, \bibinfo
  {author} {\bibfnamefont {J.}~\bibnamefont {Chen}}, \bibinfo {author}
  {\bibfnamefont {Y.}~\bibnamefont {Gao}}, \ and\ \bibinfo {author}
  {\bibfnamefont {C.~B.}\ \bibnamefont {Murray}},\ }\href {\doibase
  10.1021/nl304478h} {\bibfield  {journal} {\bibinfo  {journal} {Nano Lett.}\
  }\textbf {\bibinfo {volume} {13}},\ \bibinfo {pages} {765} (\bibinfo {year}
  {2013})}\BibitemShut {NoStop}%
\bibitem [{\citenamefont {Gorelikov}\ and\ \citenamefont
  {Matsuura}(2008)}]{Gorelikov2008}%
  \BibitemOpen
  \bibfield  {author} {\bibinfo {author} {\bibfnamefont {I.}~\bibnamefont
  {Gorelikov}}\ and\ \bibinfo {author} {\bibfnamefont {N.}~\bibnamefont
  {Matsuura}},\ }\href {\doibase 10.1021/nl0727415} {\bibfield  {journal}
  {\bibinfo  {journal} {Nano Lett.}\ }\textbf {\bibinfo {volume} {8}},\
  \bibinfo {pages} {369} (\bibinfo {year} {2008})}\BibitemShut {NoStop}%
\bibitem [{\citenamefont {Peng}\ \emph {et~al.}(2013)\citenamefont {Peng},
  \citenamefont {Smallenburg}, \citenamefont {Imhof}, \citenamefont
  {Dijkstra},\ and\ \citenamefont {van Blaaderen}}]{Peng2013}%
  \BibitemOpen
  \bibfield  {author} {\bibinfo {author} {\bibfnamefont {B.}~\bibnamefont
  {Peng}}, \bibinfo {author} {\bibfnamefont {F.}~\bibnamefont {Smallenburg}},
  \bibinfo {author} {\bibfnamefont {A.}~\bibnamefont {Imhof}}, \bibinfo
  {author} {\bibfnamefont {M.}~\bibnamefont {Dijkstra}}, \ and\ \bibinfo
  {author} {\bibfnamefont {A.}~\bibnamefont {van Blaaderen}},\ }\href {\doibase
  10.1002/anie.201301520} {\bibfield  {journal} {\bibinfo  {journal} {Angew.
  Chem., Int. Ed.}\ }\textbf {\bibinfo {volume} {52}},\ \bibinfo {pages} {6709}
  (\bibinfo {year} {2013})}\BibitemShut {NoStop}%
\bibitem [{\citenamefont {de~Nijs}\ \emph {et~al.}(2014)\citenamefont
  {de~Nijs}, \citenamefont {Dussi}, \citenamefont {Smallenburg}, \citenamefont
  {Meeldijk}, \citenamefont {Groenendijk}, \citenamefont {Filion},
  \citenamefont {Imhof}, \citenamefont {Dijkstra},\ and\ \citenamefont {van
  Blaaderen}}]{deNijs2014}%
  \BibitemOpen
  \bibfield  {author} {\bibinfo {author} {\bibfnamefont {B.}~\bibnamefont
  {de~Nijs}}, \bibinfo {author} {\bibfnamefont {S.}~\bibnamefont {Dussi}},
  \bibinfo {author} {\bibfnamefont {F.}~\bibnamefont {Smallenburg}}, \bibinfo
  {author} {\bibfnamefont {J.~D.}\ \bibnamefont {Meeldijk}}, \bibinfo {author}
  {\bibfnamefont {D.~J.}\ \bibnamefont {Groenendijk}}, \bibinfo {author}
  {\bibfnamefont {L.}~\bibnamefont {Filion}}, \bibinfo {author} {\bibfnamefont
  {A.}~\bibnamefont {Imhof}}, \bibinfo {author} {\bibfnamefont
  {D.}~\bibnamefont {Dijkstra}}, \ and\ \bibinfo {author} {\bibfnamefont
  {A.}~\bibnamefont {van Blaaderen}},\ }\href@noop {} {\  (\bibinfo {year}
  {2014})}\BibitemShut {NoStop}%
\bibitem [{\citenamefont {Kremer}\ \emph {et~al.}(1996)\citenamefont {Kremer},
  \citenamefont {Mastronarde},\ and\ \citenamefont {McIntosh}}]{Kremer1996}%
  \BibitemOpen
  \bibfield  {author} {\bibinfo {author} {\bibfnamefont {J.~R.}\ \bibnamefont
  {Kremer}}, \bibinfo {author} {\bibfnamefont {D.~N.}\ \bibnamefont
  {Mastronarde}}, \ and\ \bibinfo {author} {\bibfnamefont {J.~R.}\ \bibnamefont
  {McIntosh}},\ }\href {\doibase 10.1006/jsbi.1996.0013} {\bibfield  {journal}
  {\bibinfo  {journal} {J. Struct. Biol.}\ }\textbf {\bibinfo {volume} {116}},\
  \bibinfo {pages} {71} (\bibinfo {year} {1996})}\BibitemShut {NoStop}%
\bibitem [{\citenamefont {Mastronarde}(1997)}]{Mastronarde1997}%
  \BibitemOpen
  \bibfield  {author} {\bibinfo {author} {\bibfnamefont {D.~N.}\ \bibnamefont
  {Mastronarde}},\ }\href {\doibase 10.1006/jsbi.1997.3919} {\bibfield
  {journal} {\bibinfo  {journal} {J. Struct. Biol.}\ }\textbf {\bibinfo
  {volume} {120}},\ \bibinfo {pages} {343} (\bibinfo {year}
  {1997})}\BibitemShut {NoStop}%
\bibitem [{\citenamefont {Grzelczak}\ \emph {et~al.}(2012)\citenamefont
  {Grzelczak}, \citenamefont {S\'{a}nchez-Iglesias}, \citenamefont {Mezerji},
  \citenamefont {Bals}, \citenamefont {P\'{e}rez-Juste},\ and\ \citenamefont
  {Liz-Marz\'{a}n}}]{Grzelczak2012}%
  \BibitemOpen
  \bibfield  {author} {\bibinfo {author} {\bibfnamefont {M.}~\bibnamefont
  {Grzelczak}}, \bibinfo {author} {\bibfnamefont {A.}~\bibnamefont
  {S\'{a}nchez-Iglesias}}, \bibinfo {author} {\bibfnamefont {H.~H.}\
  \bibnamefont {Mezerji}}, \bibinfo {author} {\bibfnamefont {S.}~\bibnamefont
  {Bals}}, \bibinfo {author} {\bibfnamefont {J.}~\bibnamefont
  {P\'{e}rez-Juste}}, \ and\ \bibinfo {author} {\bibfnamefont {L.~M.}\
  \bibnamefont {Liz-Marz\'{a}n}},\ }\href {\doibase 10.1021/nl3021957}
  {\bibfield  {journal} {\bibinfo  {journal} {Nano Lett.}\ }\textbf {\bibinfo
  {volume} {12}},\ \bibinfo {pages} {4380} (\bibinfo {year}
  {2012})}\BibitemShut {NoStop}%
\bibitem [{\citenamefont {Wilson}(1990)}]{Wilson1990}%
  \BibitemOpen
  \bibfield  {author} {\bibinfo {author} {\bibfnamefont {T.}~\bibnamefont
  {Wilson}},\ }\href@noop {} {\emph {\bibinfo {title} {{Confocal
  Microscopy}}}}\ (\bibinfo  {publisher} {Academic Press},\ \bibinfo {year}
  {1990})\BibitemShut {NoStop}%
\bibitem [{\citenamefont {Hell}\ \emph {et~al.}(1993)\citenamefont {Hell},
  \citenamefont {Reiner}, \citenamefont {Cremer},\ and\ \citenamefont
  {Stelzer}}]{Hell1993}%
  \BibitemOpen
  \bibfield  {author} {\bibinfo {author} {\bibfnamefont {S.}~\bibnamefont
  {Hell}}, \bibinfo {author} {\bibfnamefont {G.}~\bibnamefont {Reiner}},
  \bibinfo {author} {\bibfnamefont {C.}~\bibnamefont {Cremer}}, \ and\ \bibinfo
  {author} {\bibfnamefont {E.~H.~K.}\ \bibnamefont {Stelzer}},\ }\href
  {http://onlinelibrary.wiley.com/doi/10.1111/j.1365-2818.1993.tb03315.x/abstr%
act} {\bibfield  {journal} {\bibinfo  {journal} {J. Microsc.}\ }\textbf
  {\bibinfo {volume} {169}},\ \bibinfo {pages} {391} (\bibinfo {year}
  {1993})}\BibitemShut {NoStop}%
\bibitem [{\citenamefont {Besseling}\ \emph {et~al.}()\citenamefont
  {Besseling}, \citenamefont {Hermes}, \citenamefont {Kuijk}, \citenamefont
  {Peng}, \citenamefont {Dijkstra}, \citenamefont {Imhof},\ and\ \citenamefont
  {van Blaaderen}}]{Besseling}%
  \BibitemOpen
  \bibfield  {author} {\bibinfo {author} {\bibfnamefont {T.~H.}\ \bibnamefont
  {Besseling}}, \bibinfo {author} {\bibfnamefont {M.}~\bibnamefont {Hermes}},
  \bibinfo {author} {\bibfnamefont {A.}~\bibnamefont {Kuijk}}, \bibinfo
  {author} {\bibfnamefont {B.}~\bibnamefont {Peng}}, \bibinfo {author}
  {\bibfnamefont {M.}~\bibnamefont {Dijkstra}}, \bibinfo {author}
  {\bibfnamefont {A.}~\bibnamefont {Imhof}}, \ and\ \bibinfo {author}
  {\bibfnamefont {A.}~\bibnamefont {van Blaaderen}},\ }\href@noop {} {\bibinfo
  {journal} {{in preparation}}\ }\BibitemShut {NoStop}%
\bibitem [{\citenamefont {Yunker}\ \emph {et~al.}(2011)\citenamefont {Yunker},
  \citenamefont {Chen}, \citenamefont {Zhang}, \citenamefont {Ellenbroek},
  \citenamefont {Liu},\ and\ \citenamefont {Yodh}}]{Yunker2011a}%
  \BibitemOpen
\bibfield  {journal} {  }\bibfield  {author} {\bibinfo {author} {\bibfnamefont
  {P.}~\bibnamefont {Yunker}}, \bibinfo {author} {\bibfnamefont
  {K.}~\bibnamefont {Chen}}, \bibinfo {author} {\bibfnamefont {Z.}~\bibnamefont
  {Zhang}}, \bibinfo {author} {\bibfnamefont {W.~G.}\ \bibnamefont
  {Ellenbroek}}, \bibinfo {author} {\bibfnamefont {A.~J.}\ \bibnamefont {Liu}},
  \ and\ \bibinfo {author} {\bibfnamefont {A.~G.}\ \bibnamefont {Yodh}},\
  }\href {http://pre.aps.org/abstract/PRE/v83/i1/e011403} {\bibfield  {journal}
  {\bibinfo  {journal} {Phys. Rev. E}\ }\textbf {\bibinfo {volume} {83}},\
  \bibinfo {pages} {011403} (\bibinfo {year} {2011})}\BibitemShut {NoStop}%
\bibitem [{\citenamefont {Zheng}\ \emph {et~al.}(2011)\citenamefont {Zheng},
  \citenamefont {Wang},\ and\ \citenamefont {Han}}]{Zheng2011}%
  \BibitemOpen
  \bibfield  {author} {\bibinfo {author} {\bibfnamefont {Z.}~\bibnamefont
  {Zheng}}, \bibinfo {author} {\bibfnamefont {F.}~\bibnamefont {Wang}}, \ and\
  \bibinfo {author} {\bibfnamefont {Y.}~\bibnamefont {Han}},\ }\href {\doibase
  10.1103/PhysRevLett.107.065702} {\bibfield  {journal} {\bibinfo  {journal}
  {Phys. Rev. Lett.}\ }\textbf {\bibinfo {volume} {107}},\ \bibinfo {pages}
  {065702} (\bibinfo {year} {2011})}\BibitemShut {NoStop}%
\bibitem [{\citenamefont {Mishra}\ \emph {et~al.}(2013)\citenamefont {Mishra},
  \citenamefont {Rangarajan},\ and\ \citenamefont {Ganapathy}}]{Mishra2013}%
  \BibitemOpen
  \bibfield  {author} {\bibinfo {author} {\bibfnamefont {C.~K.}\ \bibnamefont
  {Mishra}}, \bibinfo {author} {\bibfnamefont {A.}~\bibnamefont {Rangarajan}},
  \ and\ \bibinfo {author} {\bibfnamefont {R.}~\bibnamefont {Ganapathy}},\
  }\href {\doibase 10.1103/PhysRevLett.110.188301} {\bibfield  {journal}
  {\bibinfo  {journal} {Phys. Rev. Lett.}\ }\textbf {\bibinfo {volume} {110}},\
  \bibinfo {pages} {188301} (\bibinfo {year} {2013})}\BibitemShut {NoStop}%
\bibitem [{\citenamefont {Edmond}\ \emph {et~al.}(2012)\citenamefont {Edmond},
  \citenamefont {Elsesser}, \citenamefont {Hunter}, \citenamefont {Pine},\ and\
  \citenamefont {Weeks}}]{Edmond2012}%
  \BibitemOpen
  \bibfield  {author} {\bibinfo {author} {\bibfnamefont {K.~V.}\ \bibnamefont
  {Edmond}}, \bibinfo {author} {\bibfnamefont {M.~T.}\ \bibnamefont
  {Elsesser}}, \bibinfo {author} {\bibfnamefont {G.~L.}\ \bibnamefont
  {Hunter}}, \bibinfo {author} {\bibfnamefont {D.~J.}\ \bibnamefont {Pine}}, \
  and\ \bibinfo {author} {\bibfnamefont {E.~R.}\ \bibnamefont {Weeks}},\ }\href
  {\doibase 10.1073/pnas.1203328109} {\bibfield  {journal} {\bibinfo  {journal}
  {Proc. Natl. Acad. Sci. U.S.A.}\ }\textbf {\bibinfo {volume} {109}},\
  \bibinfo {pages} {17891} (\bibinfo {year} {2012})}\BibitemShut {NoStop}%
\bibitem [{\citenamefont {Ball}(2013)}]{Ball2013}%
  \BibitemOpen
  \bibfield  {author} {\bibinfo {author} {\bibfnamefont {P.}~\bibnamefont
  {Ball}},\ }\href {\doibase 10.1038/nmat3720} {\bibfield  {journal} {\bibinfo
  {journal} {Nat. Mater.}\ }\textbf {\bibinfo {volume} {12}},\ \bibinfo {pages}
  {696} (\bibinfo {year} {2013})}\BibitemShut {NoStop}%
\bibitem [{\citenamefont {Bates}\ and\ \citenamefont
  {Frenkel}(1998)}]{Bates1998}%
  \BibitemOpen
  \bibfield  {author} {\bibinfo {author} {\bibfnamefont {M.~A.}\ \bibnamefont
  {Bates}}\ and\ \bibinfo {author} {\bibfnamefont {D.}~\bibnamefont
  {Frenkel}},\ }\href {\doibase 10.1063/1.477248} {\bibfield  {journal}
  {\bibinfo  {journal} {J. Chem. Phys.}\ }\textbf {\bibinfo {volume} {109}},\
  \bibinfo {pages} {6193} (\bibinfo {year} {1998})}\BibitemShut {NoStop}%
\end{thebibliography}%
\end{document}